\documentclass[aps,pra,preprint,groupedaddress,showkeys,floatfix,showpacs]{revtex4}
\usepackage[dvips,dvipdfmx]{graphicx}
\usepackage{bm}
\usepackage{setspace}
\usepackage{mathtools}
\usepackage{amsmath}

\begin{document}

\title{Circular dichroism in angular distribution of electron-hydrogen scattering in a two-color bicircular laser field
}

\author{Gabriela Buica}
\email{buica@spacescience.ro}
\affiliation{
Institute of Space Science, P.O. Box MG-36, Ro 77125,
Bucharest-M\u{a}gurele, Romania}

\begin{abstract}

We study the origin of dichroic effects in elastic scattering
of high energy electrons by hydrogen atoms
 in the presence of a two-color bicircular laser field of commensurate frequencies,
 in the domain of moderate  intensities  below $10$ TW/cm$^2$.
We use a semiperturbative approach in which
the interaction of the hydrogen atom  with the laser field is
 treated in second-order perturbation theory,
while the interaction of the projectile electron with the laser field
is described by Gordon-Volkov wave functions.
An analytical formula of circular dichroism in the angular distribution of scattered electrons
 is derived in the weak-field domain for a two-color
laser field that is a combination of the fundamental and its third harmonic.
A comparison between the two-photon differential cross sections for
two-color co- and counterrotating circularly polarized laser fields is made
 and the effect of the intensity ratio of the monochromatic
 field components on the circular dichroism is investigated.
The dichroic effect in the angular distribution of scattered electrons
for two-photon absorption is analyzed as a function of
 the scattering and azimuthal angles.
We show that the two-color bicircular laser field can induce a strong circular dichroism
in the angular distribution of scattered electrons  at small scattering angles
where the atomic dressing effects are important, as well at larger scattering angles.
At small scattering  angles we demonstrate that the dichroic effect for two-photon transitions
 can be  predicted under the following conditions: the scattering process
 is treated in fist-order Born approximation and
the dressing of the atomic states by the laser field
is carried out at least in first-order time-dependent perturbation theory.

\end{abstract}
\date{\today}
\pacs{34.80.Qb, 34.50.Rk, 32.80.Wr}
\keywords{circular dichroism,bicircular laser field,free-free transitions,elastic scattering,atomic dressing}
\maketitle

\section{INTRODUCTION}
\label{I}

Dichroism, a well-known concept in classical optics, represents a property
shown by certain materials of having  absorption coefficients which depend
on the polarization of the incident radiation \cite{Chiral}.
In quantum mechanics the circular dichroism (CD) of an atom can be investigated
by considering its interaction with circularly polarized (CP) radiation.
The study of CD in laser-induced or laser-assisted atomic processes
has attracted an increasing theoretical as well experimental interest in the last 30 years
due to the possibility of investigating the dichroic properties of atomic systems
 \cite{Schonhense,Born,Klein}.
Of particular interest has become the concept of circular dichroism in angular
distribution (CDAD), which refers to the differences between the fluxes of scattered or ionized
electrons measured at definite spatial directions, caused by left and right CP laser light
 \cite{Cherepkov95}.
One of the potential interests in dichroic measurements and calculations on
laser-induced or laser-assisted atomic processes is that we can determine the relative
magnitudes and phases of the various interfering transition amplitudes by analyzing the differences between the angular distributions obtained for different polarization states of the laser field
\cite{manakov99,taieb}.
It is interesting to note that in  multiphoton ionization of unpolarized
one-electron atoms by monochromatic CP light, the cross section and the angular distribution of
 photoelectrons do not depend on the photon helicity  \cite{lamb}, i.e. there is no CD.
However, the dichroic effect exists for two-photon ionization of atoms
 \cite{manakov99} or laser-assisted electron-atom scattering  \cite{acgabielip,flegel}
with elliptically polarized fields.
A different situation occurs in two-color photoionization of unpolarized atoms
 when one XUV laser beam is CP and the other near infrared beam is linearly polarized (LP),
and CDAD can exist since  the  angular distributions of photoelectrons are slightly
different for opposite helicities of the  CP field \cite{taieb}.
Another different situation arises when two photoelectrons are simultaneously emitted upon
one-photon absorption and the angular and energy distributions of photoelectrons exhibit
a strong dependence on the photon helicity \cite{berakdar}.
Recently, the observation of CD was reported
in  two-color above threshold ionization of He atoms, in
 both differential and integral photoelectron yield \cite{mazza2014}.
In this type of experiment the He atoms are ionized by a XUV free-electron laser radiation
in the presence of an intense near infrared laser field and
the results confirmed the theoretical prediction of CD   in
 two-color multiphoton ionization of atoms  \cite{Kazansky}.

\noindent
Similarly to the laser-induced processes,  in laser-assisted electron-atom scattering
 the differential cross section (DCS) for CP light and high energy projectiles depends neither on the dynamical phase nor on the helicity of the radiation field,
in first-order Born approximation, and therefore the CD is absent \cite{acgabi2}.
However,  Manakov and coworkers \cite{manakov95} have shown that  CDAD can be predicted
in laser-assisted potential scattering of low energy electrons, provided
the CP laser has low frequency and low intensity, and if the scattering amplitude
is evaluated to higher orders in the Born series.
In the last two decades theoretical studies of dichroic effects involving
 \textit{monochromatic} electromagnetic fields,
with various combinations of linear and circular polarizations,  were performed
 for laser-assisted electron-hydrogen scattering by Cionga and coworkers
\cite{acgabiopt}, in which  the  dressing of atomic states is taken
into account in second-order of time-dependent perturbation theory (TDPT).
Very recently, the study of laser-assisted electron-atom scattering
has attracted considerable attention,
especially because of the progress of experimental techniques \cite{musa,Harak,Kanya,Kanya2}.
Despite the theoretical and experimental studies
 the roles of the initial and final atomic states, the intermediate resonances,
and the laser field parameters are still subjects of discussion.
Detailed reports on the laser-assisted electron-atom collisions can be found
in several review papers \cite{mason,ehl1998} and books \cite{bransden,joa2012},
and references therein.
A \textit{two-color} bicircular electromagnetic field,
which has lately attracted a lot of interest,
 consists of a superposition of two CP  fields
of different photon energies, which rotate in the same plane,
with identical helicities (corotating CP fields)
 or opposite helicities (counterrotating CP fields).
The recent generation of CP  high harmonics \cite{Fleischer2014}
 allowed the direct generation of CP soft x-ray pulses and
  has generated an increasing attention in studying different
laser-induced processes  by bicircular laser fields such as
 strong-field ionization \cite{Mancuso15}, nonsequential double ionization \cite{Mancuso16},
or laser-assisted electron-ion recombination \cite{Odzak1,Odzak}.
The physical mechanism occurring in laser-assisted or laser-induced processes
involving a two-color field  is the interference among different two-photon channels leading
to the same final state, and obviously by using CP fields the different helicities of the
photons play an important role \cite{ehl2001}.
Obviously, the symmetries of the bicircular electromagnetic field are reflected in symmetries of the DCSs \cite{gabi-pra2017}.

In contrast to past studies \cite{acgabi2,acgabiopt}, in the present manuscript
the dichroic effect is now  investigated  for the case of \textit{two-color bicircular} laser fields.
The notion of CDAD for two-color bicircular fields  refers to the
difference between DCSs of laser-assisted signals
 for the bichromatic fields with identical  or opposite helicities   \cite{taieb}.
To our knowledge, there are no other theoretical studies regarding the CD
in  laser-assisted electron-atom scattering processes in a two-color
 bicircular laser field which include the atomic dressing in second-order TDPT.
In the present contribution we shall demonstrate that is also possible
to find CDAD effects for fast electrons and small scattering angles
if the dressing of the atomic states  by the laser field is taken into account.
Our interest is to investigate the polarization effects at UV
photon energies, a domain  that has not been analyzed in detail,
where the atomic dressing effects are larger and, as a result
the dichroic effect will be larger than at lower photon energies.
Such an interest is justified because of the possibility of controlling
the atomic processes by using two-color laser fields and
by manipulating the  photon helicity of the monochromatic components of the fields
 \cite{ehl2001}.
In order to avoid any complications that could mask the dichroic effect
the photon energies are chosen such that they do not match any one-photon atomic resonance.
Since it is well known that the scattering probability decreases with increasing
 photon energy \cite{ehl1998},  we expect smaller DCSs than at  lower photon energies.
Therefore, we consider photon energies in the UV range ($3$\textendash$9$ eV)
 and   moderate laser intensities (below 10 TW/cm$^2$).
The manuscript is organized as follows.
In Sec. \ref{II} we briefly present the theoretical method used
in laser-assisted elastic electron-hydrogen scattering
to  derive the analytical formulas for the transition amplitudes
for a two-color CP laser field with different polarizations \cite{gabi-pra2017}.
As described in our previous works \cite{gabi2015,gabi2017}
a semiperturbative approach is used, in which for the interaction of
the incident and scattered  electrons with the laser field we employ
Gordon-Volkov wave functions, while the interaction of the hydrogen atom
 with the laser field is  treated in second-order TDPT.
It is well known that the analytical studies using TDPT remain very useful for
 understanding essential details of the scattering signal
due to the fact that the analytical formulas have the advantage of
giving physical insight into the scattering process.
The numerical results are discussed in Sec. \ref{III}, where
the DCSs and CDAD by co- and counterrotating  CP  fields
 are analyzed as a function of the scattering and azimuthal
angles of the projectile electron at different intensity ratios  of the monochromatic
components of the bicircular laser field.
We predict a first intuitive view of the dichroic effect in the DCS and
 provide  \textit{simple analytic formulas}, in a closed form,
of DCSs and CDAD  for a  bicircular field
which is a combination of a fundamental laser and its  third harmonic,
at low intensities or moderate intensities and large scattering angles.
Finally, the summary and conclusions are given in  Sec. \ref{IV}.
Atomic units (a.u.) are employed throughout this manuscript unless otherwise specified.

\section{Semiperturbative theory}
\label{II}

The laser-assisted \textit{elastic scattering} of electrons by  hydrogen atoms in
 a two-color laser field can be symbolically represented in the following way:
\begin{eqnarray}
e^-(E_{p},\mathbf{p})
+  {\rm H}(1s) +N_{1i} \, \gamma (\omega_1,  \bm{\varepsilon}_1)
+ N_{mi} \, \gamma (\omega_m, \bm{\varepsilon}_{m})
\to  \nonumber \\
e^-(E_{p^\prime},{\mathbf{p^\prime}})
+{\rm H}(1s) + N_{1f} \, \gamma (\omega_1,   \bm{\varepsilon}_1)
+ N_{mf} \, \gamma (\omega_m,  \bm{\varepsilon}_{m} ),
\label{process}
\end{eqnarray}

\noindent
where $ E_{p} \,(E_{p^\prime}) $ and $\mathbf{p} \,(\mathbf{p^\prime}$)  represent
the kinetic energy and  the momentum vector of the incident (scattered) projectile electron.
 Here $\gamma\,(\omega_k, \bm{\varepsilon}_k)$
denotes a photon with the energy $\omega_k $ and the unit polarization vector
$\bm{\varepsilon}_{k}$, and
$N_k=N_{ki}-N_{kf}$  is the net number of exchanged photons
between the projectile-atom system and each monochromatic component of
the two-color laser field ($k=1$ and $m$).
For commensurate energies,  $\omega_m =m \omega_1$,
the kinetic energy of the projectile electron before
and after collision obeys the following conservation relation
$ E_{p^\prime} = E_{p} + N \omega_1 $, with $N \equiv N_1  +N_m \; \omega_m/ \omega_1$.
The  two-color bicircular laser field is treated classically and  is described as a
combination of two coplanar CP electric fields,
\begin{equation}
{ \bf E} (t) =
 \frac{i}{2} \sum_{k=1,m}{ E}_{0k} \, \bm{\varepsilon}_k \, e^{-i  \omega_k  t} + \rm{ c.c.},
\label{field}
\end{equation}

\noindent
where ${E}_{0k}$ represents the amplitude of the monochromatic components of the electric field and
 $\bm{\varepsilon}_1 = ( \mathbf{e}_{j} +  i \mathbf{e}_{l} ) /\sqrt{2}$
 is the polarization vector of the first laser beam,
with $\mathbf{e}_{j}$ and  $\mathbf{e}_{l}$  unit vectors along two orthogonal directions.
The second laser beam has either the same polarization,
$ \bm{\varepsilon}_{m}  = \bm{\varepsilon}_1$, for \textit{corotating} two-color CP fields,
or   is  circularly polarized in the opposite direction,
$ \bm{\varepsilon}_{m}  =\bm{\varepsilon}_1^*$,
 for \textit{counterrotating} two-color CP fields.
We use a theoretical approach similar to the one developed in Ref. \cite{gabi-pra2017}
and, therefore, in the next sections we briefly describe the  model and approximations
used to calculate DCS and CDAD.

\subsection{Projectile electron and atomic wave functions}

We assume moderate laser intensities and high energy projectile electrons,
which imply  that the strength of the laser field is lower than the Coulomb field strength
experienced by an electron in the first Bohr orbit  and the
kinetic energy of the projectile electron is much larger than the energy of
the bound electron in the first Bohr orbit \cite{ehl1998}.
We describe the initial and final states of the projectile electron
interacting with a two-color laser field
by  Gordon-Volkov wave functions \cite{volkov},
\begin{equation}
 \chi_{ \mathbf{p}}  ({\mathbf{r}},t)={(2\pi )^{-3/2}}
        e^ { -iE_pt+i{\mathbf{p}}\cdot {\mathbf{r}}
-i  {\mathbf{p}} \cdot \bm{\alpha}_1(t) -i  {\mathbf{p}} \cdot \bm{\alpha}_m(t)
} ,
 \label{fe}
\end{equation}
where $\mathbf{r}$ is the position vector  and
$\bm{\alpha}_k(t)$, with $k=1$ and $m$, describes the classical oscillation
motion of the projectile electron in the bicircular electric fields
defined by Eq. (\ref{field}),
\begin{equation}
\bm{\alpha}_{k}(t)  = \alpha_{0k}
\left( \mathbf{e}_{j} \sin \omega_k t  \pm \, \mathbf{e}_{l} \cos \omega_k t \right)/\sqrt{2}.
\label{quiver}
\end{equation}
Here, the upper sign ($+$) is used for corotating CP fields,
while the lower sign  ($-$) is used for  counterrotating CP fields,
${\alpha_{0k}} = \sqrt{{I}_{k}}/ \omega_k^{2}$ is the quiver amplitude,  and
${I}_{k}={ E}_{0k}^2$ is the laser intensity of the monochromatic component of the two-color field.
Since the calculations  presented in this paper are made at
 moderate laser intensities,
the terms proportional to the ponderomotive energy,
$U_{p,k}= {I}_{k}/ 4 \omega_k^{2}$, are neglected
 in the Gordon-Volkov  wave function, Eq. (\ref{fe}).
At a laser intensity of $1$ TW/cm$^2$ and a photon energy
 of $3$ eV, the  ponderomotive energy is about  $0.016$ eV  and therefore
can be neglected compared to the photon and  projectile energies.

  The interaction of the hydrogen atom  with a two-color laser field at
moderate field strengths is considered within the second-order TDPT
and an approximate solution for the wave function
of an electron bound to a Coulomb potential in the presence of an
electric field is expressed as
 \begin{equation}
\Psi _{1s}\left( \mathbf{R}, t\right)  =
e^{-i { E}_{1s}t}
\left[
\psi _{1s} (\mathbf{R},t)  + \psi_{1s}^{(1)}(\mathbf{R},t)
+ \psi_{1s}^{(2)}(\mathbf{R},t)
\right]\,
,
 \label{fat}
\end{equation}
where  $\bm{R}$ denotes the position vector of the bound electron,
 ${E}_{1s}$ is the energy of the ground state, $ \psi _{1s} $
is the unperturbed  wave function of the ground state, and
  $\psi_{1s}^{(1)}$ and  $\psi_{1s}^{(2)}$ are the first- and second-order
radiative corrections to the atomic wave function.
The first-order radiative correction, $\psi_{1s}^{(1)}$, is calculated
  using the Coulomb Green's function including both  bound and continuum eigenstates,
being expressed in terms of  the  linear-response vector \cite{vf1}.
\noindent
Similarly, the second-order radiative correction to the atomic wave function, $\psi_{1s}^{(2)}$,
is expressed in terms of the quadratic response  tensors \cite{vf2}.
For a two-color laser field, the explicit forms of the first-
and second-order radiative corrections
$\psi_{1s}^{(1)}$ and  $\psi_{1s}^{(2)}$ are given in Ref. \cite{gabi-pra2017}.

\subsection{The nonlinear scattering matrix and differential cross section}
\label{scm}

As mentioned before,
 we focus our study at moderate laser intensities ($I_k \le 10$ TW/cm$^2$)
and fast projectile electrons ($E_{p} \ge 100$ eV) such that
the interaction between the projectile electron and  hydrogen
atom is well treated within the first-order Born approximation
in the static scattering potential $V(r, R)=-1/r+ 1/{|\bf{R}-\bf{r}|}$.
We employ a semiperturbative approach of the
scattering process similar to that proposed by Byron and Joachain \cite{b-j},
in which the exchange scattering can be safely neglected
and the scattering matrix \cite{massey} is calculated  at  high projectile energies as
\begin{equation}
S_{fi} = -i \int_{-\infty}^{+\infty} dt
\langle \chi_{ \mathbf{p^\prime}}( \mathbf{r},t)
\Psi_{1s}(\mathbf{R},t)  		 |V(r,R) |
 	      {\chi}_{ \mathbf{p}}( \mathbf{r}, t)
\Psi_{1s}(\mathbf{R},t)
			\rangle .
\,
\label{sm}
\end{equation}
\noindent
 $\chi_{\mathbf{p}}$  and $\chi_{\mathbf{p^\prime}}$ are given by Eq. (\ref{fe})
and  represent the initial and final
Gordon-Volkov wave functions  of the projectile electron  embedded in the two-color laser field,
 whereas $\Psi_{1s} $  represents the  wave function of the bound electron
interacting with the two-color  laser field and is  calculated from Eq. (\ref{fat}).
Using the  Jacobi-Anger identity \cite{Watson},
$
e^{i a  \sin \omega  t} =\sum_{N} J_N(a) e^{i N \omega t},
$
we expand the field dependent part of the Gordon-Volkov wave functions,
$\chi_{\mathbf{p}}$  and $\chi_{\mathbf{p^\prime}}$, in the scattering matrix
 in terms of the phase-dependent generalized Bessel functions,
 $ B_N$, \cite{varro}, as
\noindent
\begin{equation}
\exp{[-i  \bm{\alpha}_1(t)  \cdot \mathbf{q} -i  \bm{\alpha}_m(t)  \cdot \mathbf{q}]} =
\sum_{N=-\infty}^{+\infty} B_N({\cal R}_{1},{\cal R}_{m};\phi_{1},\phi_{m})
 e^{-i N \omega_1 t +i N \phi_{1}},
\label{bgv}
\end{equation}
\noindent
where
\begin{equation}
 B_N({\cal R}_{1},{\cal R}_{m};\phi_{1},\phi_{m})
=\sum_{l=-\infty}^{+\infty}
 J_{N-ml}({\cal R}_{1})  J_{l}({\cal R}_{m}) e^{-i  l(m\phi_{1}- \phi_{m})},
\label{bn}
\end{equation}
\noindent
in which $ J_{N-ml}({\cal R}_{1}) $ and $ J_{l}({\cal R}_{m})$ are ordinary
Bessel functions of the first kind.
The argument of the Bessel functions of the first kind is defined by
 $ {\cal R}_{k}= \alpha_{0k}|\bm{\varepsilon}_k \cdot \mathbf{q} |$, ($k=1$ and $m$),
 and $\phi_{k}$ is the dynamical phase calculated as
$ e^{i \phi_{k}} =  \bm{\varepsilon}_k\cdot \mathbf{q}
/ |\bm{\varepsilon}_k\cdot \mathbf{q} |$,
where $\mathbf{q}$ denotes the momentum transfer vector of projectile during scattering,
i.e., $ \mathbf{q}= \mathbf{p} - \mathbf{p^\prime}$.
We note that for a monochromatic CP field with the polarization unit vector
$\bm{\varepsilon}_k= ( \mathbf{e}_{j} +  i \mathbf{e}_{l} ) /\sqrt{2}$,
we obtain
$ {\cal R}_{k}= \alpha_{0k}\sqrt{(\mathbf{e}_j\cdot \mathbf{q})^2
+(\mathbf{e}_l\cdot \mathbf{q})^2}/\sqrt{2}$ and
$\phi_{k}=\arctan{(\mathbf{e}_l\cdot \mathbf{q})/(\mathbf{e}_j\cdot \mathbf{q})} +s\pi$,
 where $s$ is an integer.
Clearly, a change of helicity of the CP field,  i.e., $\bm{\varepsilon}_{k} \to \bm{\varepsilon}^*_{k}$,
leads to a change of the sign of the dynamical phase, $\phi_{k} \to -\phi_{k}$.
By replacing Eqs. (\ref{fe}), (\ref{fat}), and  (\ref{bgv}) in
 Eq. (\ref{sm}) we obtain, after integrating over time and over
 projectile coordinate, the scattering matrix for
 elastic electron-hydrogen collisions in a two-color laser field,

\begin{equation}
S_{fi} =
- 2\pi i \sum_{N=-\infty}^{+\infty}T_{fi,N} \,
\delta( E_{p^\prime}  -   E_{p}  - N \omega_1) \,,
\label{smt}
\end{equation}
\noindent
where in the $\delta$ Dirac function the kinetic energy of the projectile
is modified by $ N\omega_1 $.
Hence, the energy spectrum of the scattered electron consists of an elastic line $N=0$
 ($E_{p^\prime}=E_{p} $ and $N_{ki}=N_{kf}$),
 and of a number of sidebands corresponding to the positive and negative values of $N$.
The total nonlinear transition amplitude,  $ T_{fi,N} $,
for the elastic scattering  process  is expressed as a sum of three terms

\begin{equation}
T_{fi,N} = T^{(0)}_{N} + T^{(1)}_{N}+ T^{(2)}_{N}
,\label{tgen}
\end{equation}
\noindent
and the nonlinear DCS of the  scattered  electrons is calculated as
\begin{equation}
\frac{d{\sigma}_{N}}{d\Omega_{p'}} =
 {(2\pi)}^4  \frac{{p^\prime}}{p}  {| T^{(0)}_{N} + T^{(1)}_{N}+ T^{(2)}_{N} |}^2
,\label{dcs}
\end{equation}
\noindent
where the final momentum of the projectile is given by
${p^\prime}  = {( p^{2} + 2N \omega_1 ) }^{1/2}$.

The derivation of the transition amplitudes $T^{(i)}_{N}$, ($i=0,1,2$),
is briefly described in what follows.
The first term  on the right-hand side of  the total transition amplitude
 Eq. (\ref{tgen}),  $T^{(0)}_{N}$, is the elastic transition amplitude
due to projectile electron contribution in which the atomic dressing is neglected,

\begin{equation}
T^{(0)}_{N} =
B_N ({\cal R}_{1},{\cal R}_{m};\phi_{1},\phi_{m}) F(\mathbf{q}) ,
\label{tn0}
\end{equation}

\noindent
where the  atomic form factor is given by
$ F(\mathbf{q}) = {(2 \pi^2 q^2)}^{-1}\langle\psi_{1s}|
	 e^{i  \mathbf{q} \cdot \mathbf{R} } - 1
|\psi_{1s}\rangle
\label{ff}$.
After  performing the radial integration in the atomic form factor,
we obtain the electronic transition amplitude
\begin{equation}
T^{(0)}_{N} =- \frac{1}{(2\pi)^2}
 B_N ({\cal R}_{1},{\cal R}_{m};\phi_{1},\phi_{m})\, f_{el}^{B_1}(q),
\label{tne}
\end{equation}
in which  $ f_{el}^{B_1}(q) ={ 2(q^{2}+8) }/{(q^2+4)^2}$
is   the first-order Born approximation of the scattering amplitude for the
elastic scattering process in the absence of the laser field.
The laser field dependence  is contained in  the arguments of the
 generalized  Bessel function only $B_N({\cal R}_{1},{\cal R}_{m};\phi_{1},\phi_{m}) $.

The second term on the right-hand side of  Eq. (\ref{tgen}), $T^{(1)}_{N}$,
represents the first-order atomic transition amplitude and
occurs due to modification  of the atomic ground state by the
two-color laser field (\textit{atomic dressing})  which is  described
 by  the first-order radiative correction, $\psi^{(1)}_{1s}(\mathbf{R},t)$.
 After some calculation, the first-order atomic transition amplitude can be written as
\begin{equation}
T^{(1)}_{N} = -\sum_{k=1,m} \frac{\alpha_{0k} \omega_k}{2}
\left[ B_{N-k}\;
{\cal M}_{at}^{(1)} ( \omega_{k} ,\mathbf{q}) e^{-ik\phi_{1}}+
B_{N+k}\;
 {\cal M}_{at}^{(1)} ( -\omega_{k}, \mathbf{q} ) e^{ik\phi_{1}} \right],
\label{tn1}
\end{equation}
\noindent
where  $ {\cal M}_{at}^{(1)}(\omega_k,\mathbf{q})$ denote specific
first-order atomic transition matrix elements related to one-photon absorption,
whereas the transition matrix elements $ {\cal M}_{at}^{(1)} ( -\omega_{k}, \mathbf{q} ) $
 are related to one-photon emission \cite{gabi-pra2017}.
Obviously, in Eq. (\ref{tn1}) only one photon is exchanged (emitted or absorbed)
between the two-color laser  and  the bound electron,
while the remaining $N \pm 1$ photons are exchanged between the two-color laser
and the projectile electron.
 For the sake of simplicity, the arguments of the generalized Bessel functions
are dropped off in Eq. (\ref{tn1})  and throughout this paper.
By performing the radial integrals  in Eq. (\ref{tn1}) we find
 the  first-order atomic transition amplitude, $T^{(1)}_{N}$, as,
\begin{eqnarray}
T^{(1)}_{N} &=&
\sum_{k=1,m} \frac{\alpha_{0k} \omega_k}{4{\pi}^{2}q^2 }
\left[
  (\bm{\varepsilon}_k  \cdot \hat {\mathbf{q}}) B_{N-k} {\cal J}_{101}(\omega_{k},q)
e^{-i k\phi_1} \right.
\nonumber \\ && \left.
- (\bm{\varepsilon}_k^*\cdot \hat {\mathbf{q}}) B_{N+k} {\cal J}_{101}(\omega_{k},q)
e^{ik\phi_1}\right],
\label{tni}
\end{eqnarray}
where the  radial integral ${\cal J}_{101}(\omega_{k},q) $
 is analytically calculated  as series of hypergeometric functions \cite{acgabi2,gabi2017}
and $\hat{\mathbf{q}} = \mathbf{q} /| \mathbf{q}|$
represents a unit vector that defines the direction of the momentum transfer vector.

The last term on the right-hand side of the transition amplitude
 Eq. (\ref{tgen}), $T^{(2)}_{N}$, represents the second-order atomic transition amplitude
and occurs due to  alteration of the atomic ground state by the two-color laser field
which is described by  the second-order radiative correction, $\psi^{(2)}_{1s}(\mathbf{R},t)$,
calculated in Ref. \cite{gabi-pra2017}.
After some algebra the second-order atomic transition amplitude, $T^{(2)}_{N}$, is expressed as
\begin{eqnarray}
T^{(2)}_{N} &=& \sum_{k=1,m}\frac{ \alpha_{0k}^2 \omega_k ^2 }{4}
    \left\{
 B_{N-2k}         {\cal M}_{at}^{(2)}
            ( \omega_k,  \mathbf{q})  e^{-2ik\phi_{1}}
+
B_{N+2k}        {\cal M}_{at}^{(2)}
            ( - \omega_k, \mathbf{q} ) e^{2ik\phi_{1}} \right.
 \nonumber \\
    &&     \left.
+ B_{N}
 \left[\widetilde {\cal M}_{at}^{(2)} ( { E}_{1s},\omega_k )
               +\widetilde {\cal M}_{at}^{(2)}  ( { E}_{1s},-\omega_k )\right]
\right\}
\nonumber \\
&&+ \frac{\alpha_{01} \alpha_{0m} \omega_1 \omega_m }{4}
   \sum_{l,j=\pm 1}
 B_{N-l-jm}
        {\cal N}_{at}^{(2)}
       ( l\omega_1, j\omega_m, \mathbf{q}) e^{-i(l+jm)\phi_{1}}
.\label{t2}
\end{eqnarray}
\noindent
Obviously, in Eq. (\ref{t2}) only two photons are exchanged (absorbed, emitted,
or absorbed and emitted) between the two-color laser field and  the bound electron.
The specific second-order atomic transition matrix elements,
$ {\cal M}_{at}^{(2)}$, $\widetilde {\cal M}_{at}^{(2)}$,
and $ {\cal N}_{at}^{(2)}$ are connected to two-photon exchange
and are calculated  in Refs. \cite{acgabi99,acgabi2}.
The atomic transition matrix elements
${\cal M}_{at}^{(2)}(\omega_{k}, \mathbf{q} )  $ and
$ {\cal M}_{at}^{(2)} (-\omega_{k}, \mathbf{q} ) $
are related to absorption and emission of two identical photons of energy
 $\omega_k$ and complex polarization $\bm{\varepsilon}_{k}$, respectively.
The second type of atomic transition matrix elements
 $\widetilde {\cal M}_{at}^{(2)}( \omega_k,\mathbf{q} )$
 describe the absorption  followed by emission  of the same photon,
whereas $\widetilde {\cal M}_{at}^{(2)}( -\omega_k,\mathbf{q} )$
 describe the emission followed by absorption  of the same photon.
 The third type of atomic transition matrix elements
${\cal N}_{at}^{(2)}( \omega_1, \omega_m, \mathbf{q}) $ and
 ${\cal N}_{at}^{(2)} (- \omega_1, -\omega_m, \mathbf{q}) $
describe the  absorption and emission   of two different photons of
energies  $\omega_1$ and $\omega_m$, respectively.
Similarly, ${\cal N}_{at}^{(2)} (- \omega_1, \omega_m, \mathbf{q}) $
are related to  emission of one photon of energy  $\omega_1$ and
 absorption of  one photon of energy  $\omega_m$.
The  analytic expressions of the second-order atomic transition matrix elements
for two \textit{identical} photons are given by
\begin{equation}
{\cal M}_{at}^{(2)}(\omega_k,{\mathbf{q}})=
\frac{(\bm{\varepsilon}_k \cdot \hat {\mathbf{q}})^2}{2 \pi^2 q^2}
 {\cal Q}(\omega_k,{q})
+\frac{ \bm{\varepsilon}_k^2}{2 \pi^2 q^2}
 {\cal P}(\omega_k,{q})
,
\label{m2}
\end{equation}
and
\begin{equation}
\widetilde {\cal M}_{at}^{(2)}(\omega_k,{\mathbf{q}})=
\frac{|\bm{\varepsilon}_k \cdot \hat {\mathbf{q}}|^2}{2 \pi^2 q^2}
 \widetilde{\cal Q}(\omega_k,{q})+
\frac{1}{2 \pi^2 q^2}
 \widetilde{\cal P}(\omega_k,{q})
,
\label{m2t}
\end{equation}
\noindent
  where the specific  expressions of the polarization-invariant atomic radial integrals
$ {\cal P}$ and ${\cal Q}$ for two-photon processes
depend on the photon energies and the amplitude of the momentum transfer vector
 of the projectile electron,
and are obtained  as series of hypergeometric functions in Ref.  \cite{acgabi2}.
We note that  $\bm{\varepsilon}_k^2=1$ for a LP field, while
$\bm{\varepsilon}_k^2=0$ for a CP field
and therefore, the second term in the right-hand side of Eq. (\ref{m2}) vanishes.
\noindent
For the exchange of two  \textit{different} photons a general form of
 the second-order atomic transition matrix elements was derived  \cite{acgabi99} as
\begin{equation}
{\cal N}_{at}^{(2)}(\omega_j,\omega_l,{\mathbf{q}})=
\frac{(\bm{\varepsilon}_j\cdot \hat {\mathbf{q}})
(\bm{\varepsilon}_l \cdot \hat {\mathbf{q}})}
{2 \pi^2 q^2}  {\cal Q}^{\prime}(\omega_j,\omega_l,{q})
+\frac{\bm{\varepsilon}_j  \cdot \bm{\varepsilon}_l}{2 \pi^2 q^2}
 {\cal P}^{\prime}(\omega_j,\omega_l,{q}).
\label{n2gen}
\end{equation}
\noindent
Equation (\ref{n2gen}) is symmetric and has a structure
that explicitly contains only the scalar products of polarization and momentum transfer vectors.
The general structure of Eq. (\ref{n2gen})  is also similar to other processes, such as
the elastic scattering of photons by hydrogen atoms \cite{gavrila1970},
 two-photon bremsstrahlung  \cite{gavrila},
 elastic x-ray scattering by ground-state atoms \cite{manakov},
two-photon ionization of hydrogen \cite{taieb,fifirig2000}, or
two-photon double ionization  \cite{starace},
with the unit vector $\hat {\mathbf{q}}$ replaced
by  vectors which are specific to each particular process.
The specific  expressions of the polarization-invariant atomic radial integrals
for two different photons, $ {\cal P}^{\prime}$ and ${\cal Q}^{\prime}$,
 depend on the photon energies
$ \omega_{j}$ and $ \omega_{l}$, and the momentum transfer $q$
\cite{acgabi99}.
The following  the changes are made $ \omega_{k} \to -\omega_{k}$ and
 $\bm{\varepsilon}_{k} \to \bm{\varepsilon}_{k}^*$, if the photon $k$ is emitted
($k=j $ and $l$).
Clearly, for the absorption of two distinct photons
with identical circular polarizations $\bm{\varepsilon}_{j}=\bm{\varepsilon}_{l}$
(corotating CP fields) the second term on the right-hand side of Eq. (\ref{n2gen}) vanishes,
while for opposite polarizations $\bm{\varepsilon}_{l}=\bm{\varepsilon}_{j}^*$
(counterrotating CP fields)
we have $\bm{\varepsilon}_{j} \cdot \bm{\varepsilon}_{l}=1$.

Finally, the CDAD is defined as the difference between the laser-assisted DCSs
 for co- and counterrotating two-color CP fields
\begin{equation}
\Delta_{CDAD}( \theta,\varphi )=
   \frac{d\sigma_{N}^{++}}{d\Omega_{p'}}( \theta,\varphi )
-  \frac{d\sigma_{N}^{+-}}{d\Omega_{p'}}( \theta,\varphi ),
\label{CD}
\end{equation}
where the superscript $++$ indicates that both monochromatic components of the field
 have  identical polarizations (corotating fields), while superscript $+-$
indicates that the monochromatic components have opposite polarizations (counterrotating fields).
 It is interesting to note that, the explicit or implicit presence of
the dynamical phase factors $ e^{i \phi_1}$ and $ e^{ i \phi_m}$
 in  the electronic as well the first-and second-order atomic transitions amplitudes,
 Eqs. (\ref{tne}), (\ref{tni}),  and (\ref{t2}),  can give  different interference terms
 in the expression of DCS  for corotating  in comparison to counterrotating CP fields.
\noindent
For practical reasons the circular dichroism in angular distribution can be
better visualized through a relative CDAD, defined as the ratio between the
 difference of DCSs for corotating and counterrotating fields and the sum of these DCSs,

\begin{equation}
R(\theta,\phi) =
\Delta_{CDAD}( \theta,\varphi )
\left[ \frac{d\sigma_{N}^{++}}{d\Omega_{p'}}( \theta,\varphi )
+  \frac{d\sigma_{N}^{+-}}{d\Omega_{p'}}( \theta,\varphi )\right]^{-1}.
\label{RCDAD}
\end{equation}
\noindent
Obviously, the relative CDAD defined in the above equation approaches  values of
$+ 1$ or $- 1$ in those cases where one of the DCSs,
${d\sigma_{N}^{++}}/{d\Omega_{p'}}$ or
$ {d\sigma_{N}^{+-}}/{d\Omega_{p'}}$,
 is almost equal to zero, whereas  the relative CDAD  approaches the value
zero in those cases where
${d\sigma_{N}^{++}}/{d\Omega_{p'}} \simeq {d\sigma_{N}^{+-}}/{d\Omega_{p'}}$.

\section{NUMERICAL EXAMPLES AND DISCUSSION}
\label{III}

\begin{figure}
\centering
\includegraphics[width=3.9in,angle=0]{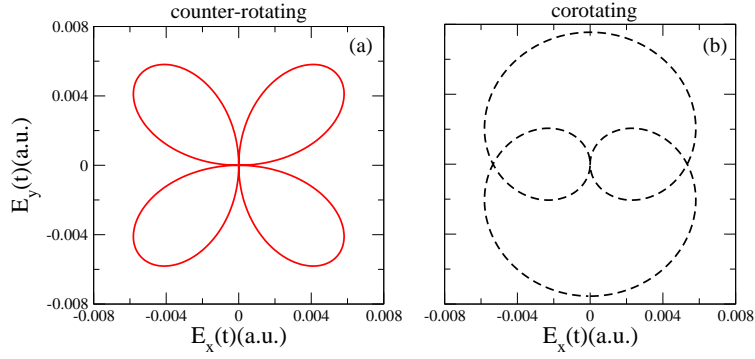}
\caption{(Color online) Parametric plots showing the Cartesian components of
the electric field vector  in the   $(x,y)$-polarization plane,
 $E_x(t)$ and $E_y(t)$,   plotted for $0 \leq t \leq T_1$,
for  two-color left- and right-handed CP fields  with
$\bm{\varepsilon}_{1}= \bm{\varepsilon}_{+}=(\mathbf{e}_x+i\mathbf{e}_y)/\sqrt{2}$ and
$\bm{\varepsilon}_{3}  = \bm{\varepsilon}_{-}$ in  panel (a),
and two-color left-handed  CP fields  with
$\bm{\varepsilon}_{1}=\bm{\varepsilon}_{3}  = \bm{\varepsilon}_{+}$ in  panel (b).
The laser field intensities are $I_1=I_3=1$ TW/cm$^2$, the fundamental photon energy is $\omega_1= 3 $ eV
while the energy of the harmonic photon  is $\omega_3=3\omega_1$.
The two-color bicircular electric field satisfies a $T_1/2$ and $T_1/4$
rotational symmetry for co- and counterrotating polarizations, respectively.
}
\label{fig1}
\end{figure}

In this section, we present numerical results for the scattering process
 described by Eq. (\ref{process}), in which two photons
are absorbed in the electron-hydrogen scattering process embedded in a two-color CP laser,
which is a combination of the fundamental and its third harmonic ($m$=3).
It is worth pointing out that Eqs.  (\ref{tne}), (\ref{tni}),  and (\ref{t2})
are applicable for arbitrary scattering configurations
and two-color laser fields with linear and/or circular polarizations.
We focus our discussion on two different polarizations where the two-color
laser beams are CP in the $(x,y)$ plane
with one laser beam propagating in the $z$-axis direction,
$ \bm{\varepsilon}_{1}  =\bm{\varepsilon}_{+}  \equiv (\mathbf{e}_x+i \mathbf{e}_y)/\sqrt{2}$
(left-handed CP),
while the other laser beam has either the same circular polarization
$ \bm{\varepsilon}_{3}  = \bm{\varepsilon}_{+}$, i.e., the \textit{corotating polarization} case,
or   is  CP in the opposite direction,
$ \bm{\varepsilon}_{3}  = \bm{\varepsilon}_{-}\equiv (\mathbf{e}_x-i \mathbf{e}_y)/\sqrt{2} $
(right-handed CP),  i.e.,   the \textit{counterrotating polarization} case.
For example, in the case of equal amplitudes
of the monochromatic field components  (${ E}_{01}={ E}_{03}$),
 the bicircular electric field defined in Eq. (\ref{field}) reduces to
\begin{equation}
{\bf E}_{+} (t) ={ E}_{01} \sqrt{2} \,
(\mathbf{e}_x \sin \omega_{+}t -\mathbf{e}_y \cos \omega_{+}t)
\cos \omega_{-}t,
\label{co}
\end{equation}
for identical circular polarizations
 ($\bm{\varepsilon}_{1}  =  \bm{\varepsilon}_{3}  = \bm{\varepsilon}_{+}$),
 and
\begin{equation}
{\bf E}_{-} (t) ={ E}_{01} \sqrt{2}\,
(\mathbf{e}_x \cos \omega_-t -\mathbf{e}_y \sin \omega_-t)  \sin \omega_+t
,\label{cn}
\end{equation}
 for opposite circular polarizations ($ \bm{\varepsilon}_{1} = \bm{\varepsilon}_{+}$
and $ \bm{\varepsilon}_{3}  = \bm{\varepsilon}_{-} $),
 where the frequencies are defined by $\omega_{\pm}=\eta \, \omega_1/2 $,
 with $\eta= 2$ for corotating CP fields and $\eta= 4$
 for counterrotating CP fields.
The bicircular electric field vectors, Eqs. (\ref{co})  and (\ref{cn}),
are invariant with respect to  translation in time by
an integer multiple of $T_1 /\eta$ and with respect to rotation in the polarization plane
 by an angle $\alpha=2\pi/\eta$ around the $z$ axis, such that
${\bf E}_{\pm}\left(t + T_1 /\eta \right) =
{\bm R }\left(2\pi /\eta \right)\, {\bf E}_{\pm}(t)$,
where ${\bm R }(\alpha)$ is a $2 \times 2$ rotation matrix with angle $\alpha$ around the $z$ axis
 and $T_1 = 2\pi/\omega_1$ is the fundamental field optical period.
\noindent
In Fig. \ref{fig1} we plot the temporal dependence of the electric field vectors,
given by   Eqs. (\ref{co})  and (\ref{cn}), in the polarization plane for two-color
 CP laser fields of equal intensities with identical polarizations
 in the  right column, and opposite polarizations in the  left column.
For counterrotating polarizations  the temporal symmetry of the electric field means that
$  {\bf E}_-(t + T_1 /4) = {\bm R } (\pi /2)  \, {\bf E}_-(t)$,
 i.e.,  the translation in time is  one-quarter of the optical cycle, $T_1/4$,
and the rotation angle in the polarization plane is $\pi/2$,
which implies a fourfold symmetry of the electric field in Fig. \ref{fig1}(a).
By contrast, for corotating polarizations the translation in time of the electric field  is $T_1/2$
and the rotation angle is $\pi$ in Fig. \ref{fig1}(b), such that
$  {\bf E}_+(t + T_1/2 ) = {\bm R } (\pi )  \, {\bf E}_+(t)$.
We expect that the symmetries of the bicircular field to be preserved
in the DCSs of the scattered electron.
The laser intensities we consider in Fig. \ref{fig1} are   $I_1=I_3=1$ TW/cm$^2$ and
 the fundamental and harmonic photon energies are $\omega_1=3$ eV and  $\omega_3=9$ eV.

\subsection{Two-photon circular dichroism in a weak bicircular laser field}
\label{IVa}

\begin{figure}
\centering
\includegraphics[width=2.9in,angle=0]{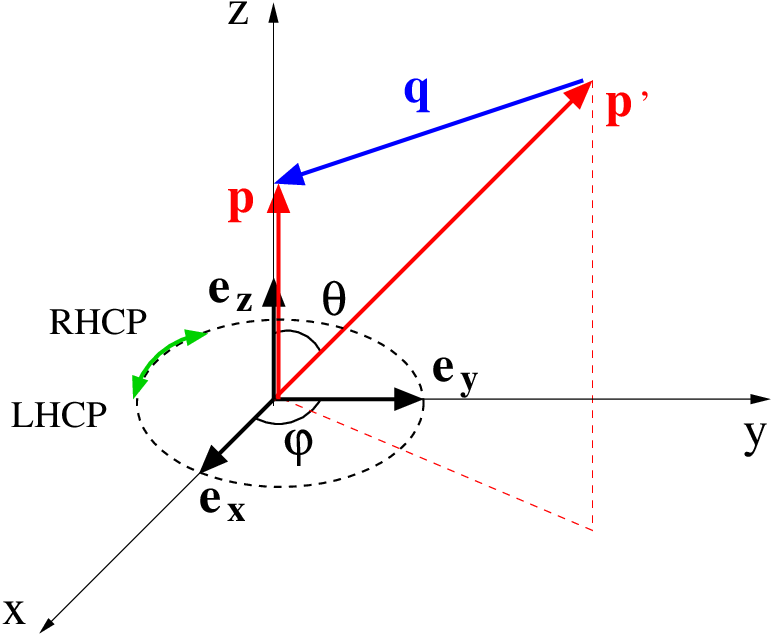}
\caption{
(Color online) The specific scattering geometry with $\mathbf{p} \parallel \mathbf{e}_z $,
where $\mathbf{p}$ and $\mathbf{p}^\prime$ are the  momentum vectors of the
incident and scattered  electron, $\theta$ is the angle between them,
$\varphi$ is the azimuthal angle, and $\mathbf{q} $ is the momentum transfer vector.
We assume that both laser beams are collinear and propagate along the $Oz$ axis.
The laser beams are CP in the $(x,y)$ plane, with the polarization vectors
$\bm{\varepsilon}_{+}=(\mathbf{e}_x+i\mathbf{e}_y)/\sqrt{2}$ left-handed-CP (LHCP) or
$\bm{\varepsilon}_{-}=(\mathbf{e}_x - i\mathbf{e}_y)/\sqrt{2}$ right-handed-CP (RHCP).
}
\label{fig2}
\end{figure}

\begin{figure}
\centering
\includegraphics[width=3in,angle=0]{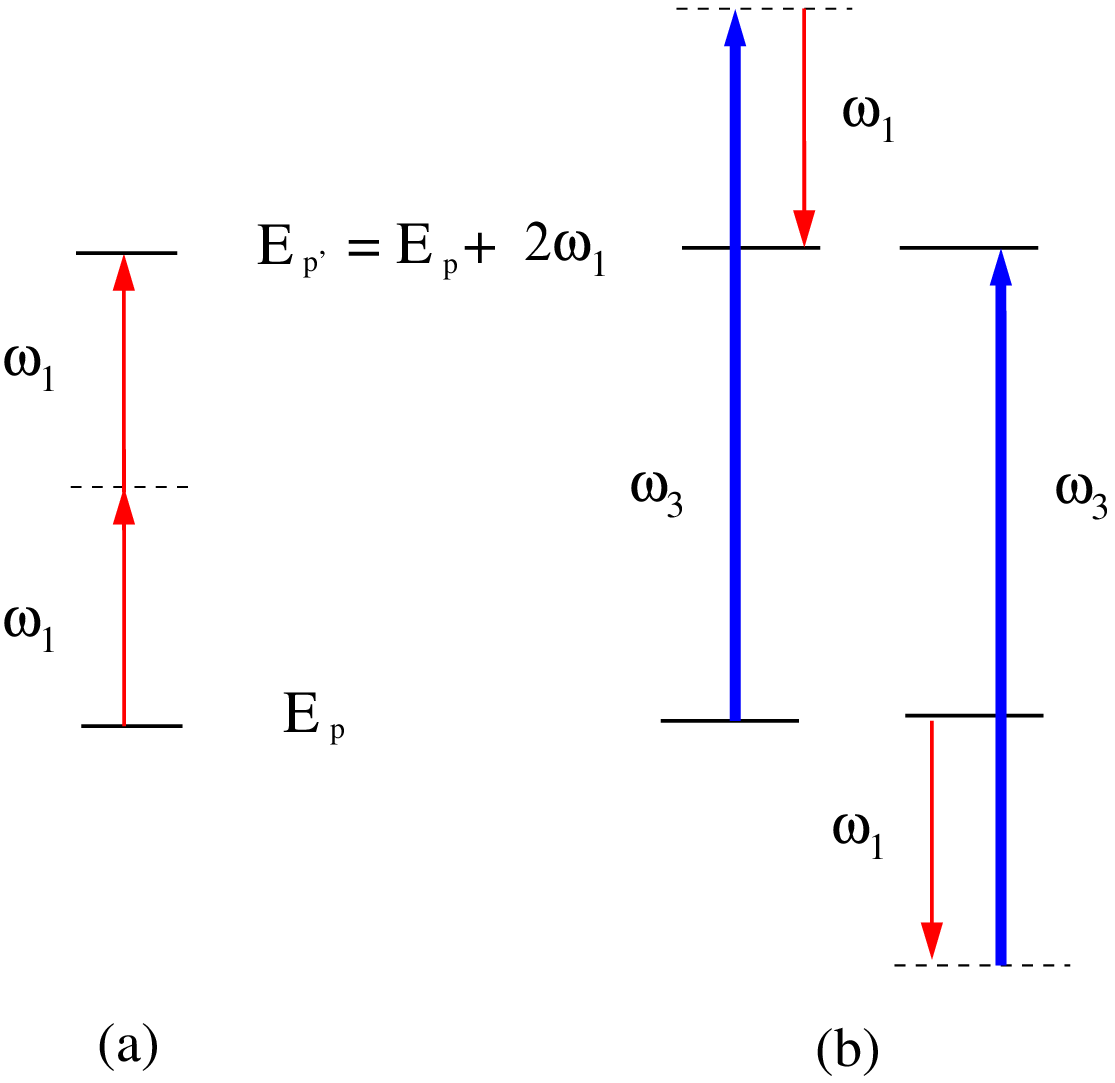}
\caption{(Color online)
 Energy diagrams schematically showing the photon channels leading
to the final energy of the projectile electron $E_{p\prime} = E_p + 2\omega_1$.
Channel (a) corresponds to  absorption of two photons of energy $\omega_1$, while channel (b)
 corresponds to absorption of one third-harmonic photon $\omega_3=3\omega_1$
and  emission of one photon $\omega_1$.
}
\label{fig3}
\end{figure}

In this section we derive simple analytical formulas
 of nonlinear DCSs and CDAD for two-photon absorption in the weak-laser-field regime,
which provide more physical insight into the dichroic effect
in the electron-hydrogen scattering process in a two-color bicircular laser field.
The total transition amplitude that includes the
first- and second-order atomic dressing can be written in a closed form that
allows us to analyze the dependence on the polarization vectors.
In what follows we consider the scattering geometry depicted in Fig. \ref{fig2}
in which the momentum vector of the incident electron  $\mathbf{p}$ is parallel
to the $z$ axis,  $\theta$ is the scattering angle between the momentum vectors
of the incident and scattered electrons ${\bf p}$ and ${\bf p^\prime}$,
and $\varphi$ is the azimuthal angle of the scattered electron.
In this scattering geometry the Cartesian components of  the momentum transfer vector
$ \mathbf{q} $ are given by
$(-p^\prime \sin \theta \cos\varphi, -p^\prime \sin \theta \sin\varphi, p -p^\prime \cos \theta )$,
 with an amplitude
$q=\sqrt{ {p^\prime}^2 +p^2-2 {p^\prime}  p \cos \theta}$
which varies in the interval $ |p^\prime -p| \leq q \leq   p^\prime +p$,
for forward ($\theta=0^{\circ}$) and backward ($\theta=180^{\circ}$) scattering,
where $p^\prime= \sqrt{p^2 +2N\omega_1}$.
Whenever the arguments of the Bessel function of the first kind are small,
 i.e., ${\cal R}_{1} \ll 1$ and ${\cal R}_{3} \ll 1$, a
condition that is satisfied either at low laser intensities or
at small scattering angles with moderate laser intensities,
the  approximate expressions of the generalized Bessel functions can be used
\cite{gabi-pra2017,Watson}.
Hence, in the limit ${\cal R}_{1(3)} \ll 1$ the  total  transition amplitude
for two-photon absorption is calculated   by keeping the second-order contributions in the fields
 and neglecting the higher powers of the fields in Eqs.  (\ref{tne}), (\ref{tni}),  and (\ref{t2}),
  as a sum of the two-photon transition amplitudes for the channels  depicted   in Fig. \ref{fig3},
\begin{eqnarray}
 T_{2}  &\simeq&
\alpha_{01}^2 |\bm{\varepsilon}_1 \cdot \mathbf{q}|^2  \,  {\cal C}_1
+
\alpha_{01}\alpha_{03}
|\bm{\varepsilon}_1^*   \cdot  {\mathbf{q}}|
|\bm{\varepsilon}_3  \cdot  {\mathbf{q}}| \, {\cal C}_2  \, e^{-i(3\phi_{1}-\phi_{3})}
 \nonumber \\ &&
+\alpha_{01}\alpha_{03}
(\bm{\varepsilon}_1^*  \cdot \bm{\varepsilon}_3) \, {\cal C}_3  \,
 e^{-2i\phi_{1}}
.
\label{t3l}
\end{eqnarray}
\noindent
The first term on the right-hand side describes absorption of two identical photons
of energy $\omega_1$,  while the rest of the terms describe absorption of one photon
 of energy $\omega_3$ and  emission of another photon of energy $\omega_1$,
 as schematically presented in Figs. \ref{fig3}(a) and \ref{fig3}(b).
The polarization-invariant amplitudes $ {\cal C}_1$, $ {\cal C}_2$,  and  $ {\cal C}_3$
depend on the momentum transfer and photon energies being defined as
\begin{eqnarray}
{\cal C}_1 & =&
\frac{1}{8\pi^2} \left[
-\frac{f_{el}^{B_1}}{4} +\frac{\omega_1}{q^3}{\cal J}_{101}(\omega_{1},q)\right],
\label{c1}\\
{\cal C}_2  & =&
\frac{1}{8\pi^2} \left[
\frac{f_{el}^{B_1}}{2} -\frac{\omega_3}{q^3}{\cal J}_{101}(\omega_{3},q)
+\frac{ \omega_1 \omega_3 }{ q^4}{\cal Q}^{\prime}(-\omega_1,\omega_3,{q})
\right],
\label{c2}\\
{\cal C}_3 & =&\frac{ \omega_1 \omega_3 }{8 \pi^2 q^2}
{\cal P}^{\prime}(-\omega_1,\omega_3,{q}).
\label{c3}
\end{eqnarray}

\noindent
In the scattering geometry shown in Fig. \ref{fig2},
 the scalar product in the argument  ${\cal R}_k$ of the generalized Bessel functions
 is given by
$ \bm{\varepsilon}_{\pm}\cdot \mathbf{q} = - {p^\prime} \sin \theta \, e^{\pm i \varphi} /\sqrt{2}$
and the dynamical phases of the  corotating laser fields are
$ \phi_{1}   =  \phi_{3}   = \pi + \varphi$,
whereas for counterrotating laser fields only the dynamical phase of the harmonic field changes as
$ \phi_{3}   =  \pi -  \varphi$.
Obviously, a change of the photon helicity in the arguments of the generalized Bessel functions
implies  a change in the sign of the azimuthal angle $\varphi$.
Therefore, in the  weak-field limit of a two-color bicircular laser,
with $\bm{\varepsilon}_3 = \bm{\varepsilon}_+$ for the corotating case
and $\bm{\varepsilon}_3 = \bm{\varepsilon}_- $ for the counterrotating case,
 we obtain the following two-photon transition amplitude,
\begin{equation}
 T_{2}  \simeq
\alpha_{01}^2 |\bm{\varepsilon}_1 \cdot \mathbf{q}|^2 \left( {\cal C}_1 \,
+
\frac{\alpha_{03}}{\alpha_{01}}  {\cal C}_2 \, e^{-i \eta \varphi}
 \right)+
 \delta_{2 \eta } \, \alpha_{01}\alpha_{03}  \, {\cal C}_3  \, e^{-2i\varphi}
 ,
\label{t3lp}
\end{equation}
\noindent
where the parameter $\eta$ is either $ 2$ for two-color  corotating CP fields
 (equal photon helicities) or  $4$ for two-color counterrotating CP fields
(opposite photon  helicities), respectively.
After substituting the scalar product, $|\bm{\varepsilon}_1 \cdot \mathbf{q}|$,
in the above equation
the corresponding two-photon  DCSs in a weak bicircular laser field
for co- and counterrotating polarizations take the following simple forms:
\begin{eqnarray}
\frac{d{\sigma}_{2}^{++}}{d\Omega_{p'}}( \theta,\varphi ) &\simeq&
 |a_1\sin^2\theta+(a_2\sin^2\theta+a_3) e^{-2i\varphi}|^2
,\label{dcsLL}
\\
\frac{d{\sigma}_{2}^{+-}}{d\Omega_{p'}}( \theta,\varphi )& \simeq&
 |a_1+a_2 e^{-4i\varphi}|^2 \sin^4\theta
,\label{dcsLR}
\end{eqnarray}
 with
$a_1= 2\pi^2  \alpha_{01}^2  {\cal C}_1 \sqrt{p^{\prime 3} /p}$,
$a_2= 2\pi^2 \alpha_{01} \alpha_{03} \,  {\cal C}_2  \sqrt{p^{\prime 3} /p} $, and
$a_3= 4\pi^2 \alpha_{01} \alpha_{03}  \, {\cal C}_3 \sqrt{p^{\prime} /p}$.
Clearly, the co- and counterrotating  DCSs, ${d{\sigma}_{2}^{++}}/{d\Omega_{p'}}$
and  $d{\sigma}_{2}^{+-}/d\Omega_{p'}$, are different
for opposite helicities of the CP harmonic laser field.
\noindent
Finally, the two-photon CDAD defined by Eq. (\ref{CD})
  is simply calculated in the weak bicircular laser field
limit  as the difference between the DCSs given by  Eqs. (\ref{dcsLL}) and  (\ref{dcsLR}),
\begin{equation}
\Delta_{CDAD}( \theta,\varphi ) \simeq
 |a_3|^2 + 2 {\rm Re}\,[a_2a_3^*\sin^2\theta
+ a_1^*\sin^2\theta(a_2\sin^2\theta+a_3) e^{-2i\varphi}
- a_1^*a_2\sin^4\theta e^{-4i\varphi} ]
.\label{dcsCDc}
\end{equation}

\noindent
Furthermore, as long as the harmonic photon energy is below the ionization threshold,
$\omega_1 < |E_{1s}|/3$,  the one- and two-photon
atomic transition matrix elements
$ {\cal M}_{at}^{(1)}, {\cal M}_{at}^{(2)}$, and $ {\cal N}_{at}^{(2)}$ are real quantities.
Therefore, CDAD can be formally expressed from Eq. (\ref{dcsCDc})
 as a function of the scattering and azimuthal angles  $ \theta $ and $\varphi$,
\begin{equation}
\Delta_{CDAD}( \theta,\varphi ) \simeq
   a_3(a_3+2a_2\sin^2\theta )
+ 2a_1\sin^2\theta(a_2\sin^2\theta+a_3)\cos (2\varphi)
- 2a_1a_2 \sin^4\theta \cos (4\varphi)
.\label{dcsCDr}
\end{equation}
Excepting the forward and backward scattering where CDAD is $\varphi$ independent,
the last two terms in the right-hand side depend on  $\cos (2 \varphi)$
and $\cos (4 \varphi)$.
  CDAD is invariant to the following transformations:
(i) $ \pi-\varphi \to \pi+\varphi $, which is equivalent to a reflection
 with respect to the $(x,z)$ plane,
and
(ii) $ \pi/2-\varphi \to \pi/2+\varphi $, which is equivalent to a reflection
 with respect to the $(y,z)$ plane.
According to Eq. (\ref{dcsCDr})  the dichroic effect can be encountered
even for rather small scattering angles  where the atomic dressing is important,
not only around the scattering angle $\theta=90^{\circ}$
as in the case of x-ray scattering  by unoriented systems \cite{manakov},
or the case of two-color two-photon photoionization
when the polarization state of one of the fields is reversed \cite{taieb}.
Obviously, as was previously mentioned, in first-order Born approximation the
dichroic effect in DCS cannot be encountered in a scattering configuration
involving only a \textit{one-color} ($\omega_1$) CP laser field  \cite{manakov95,acgabi2},
because the two-photon transition amplitude given by Eq.  (\ref{t3l})
 does not depend on the helicity of the radiation field.
However, a dichroic effect in two-photon transitions can be predicted
for a superposition of one-color LP and CP fields  under the following conditions:
 the scattering of the high-energy electrons is treated in the first-order Born approximation,
the atomic dressing effect by the laser field  is carried out in second-order TDPT, and
the transitions between the atomic bound and continuum states are energetically allowed
\cite{acgabiopt}.
In contrast to the findings of Ref. \cite{acgabiopt},  our Eq. (\ref{dcsCDr})
 shows a different regime were the dichroic effect in DCS is encountered for a
two-color bicircular laser field even if the atomic dressing is neglected
or the photons energies are below the ionization threshold, such that
$\omega _1+ \omega _3<  |E_{1s}|$.
In addition, if the atomic dressing is  small in
Eqs. (\ref{c1}), (\ref{c2}), and (\ref{c3}),
 i.e., at large scattering angles with nonresonant photon energies
or in the limit of low-photon energies,  i.e., $ E_p \gg \omega_1$,
by keeping the first-order dressing in fields
${\cal J}_{101}( \omega, q)\simeq \alpha_d \, \omega \, q$ \cite{acgabi2},
${\cal P}\simeq 0$, and ${\cal Q}\simeq 0$, we obtain
\begin{eqnarray}
a_1  & =& \frac{\alpha_{01}^2}{4 }
\sqrt\frac{p^{\prime 3}}{p }
\left( \alpha_d \frac{\omega_1^2}{q^2} - \frac{f_{el}^{B_1}}{4} \right),
 \label{c1l} \\
a_2  & =& \frac{\alpha_{01}\alpha_{03}}{4 }
\sqrt\frac{p^{\prime 3}}{p }
\left(  \frac{f_{el}^{B_1}}{2}  - \alpha_d \frac{\omega_3^2}{q^2}  \right)
, \label{c2l}\\
{a}_3 & =&0, \label{c3l}
\end{eqnarray}
where $\alpha_d$ is the dynamic dipole polarizability of the hydrogen atom in its ground state.
For high projectile energies and low photon energies, $p^{\prime }\simeq  p$ ,
the projectile momentum transfer has a simple dependence on the scattering angle, $\theta$,
 and  the following approximation formula holds for momentum transfer
 $q \simeq  p \sin (\theta/2)$.
Therefore, the CDAD is then simply expressed as
\begin{equation}
\Delta_{CDAD}( \theta,\varphi ) \simeq
 4 a_1 a_2\sin^4\theta \sin (\varphi) \sin (3\varphi)
,\label{dcsCDe}
\end{equation}
and the relative CDAD is formally calculated as
\begin{equation}
R_{CDAD}( \theta,\varphi ) \simeq
\frac{2 a_1 a_2 \sin (\varphi) \sin (3\varphi)}
{a_1^2 + a_2^2+ 2 a_1 a_2 \cos (\varphi) \cos (3\varphi)}
.\label{RCDADe}
\end{equation}
\noindent
Both  $\Delta_{CDAD}$  and $R_{CDAD}$   vanish at  azimuthal angles that are multiples of $\pi/3$.
Their absolute maxima   occur at azimuthal angles $\pi/2$ and $3\pi/2$,
 while their  minima occur at $s\pi  \pm \arccos(0.25)/2$, where $s$ is an integer.
The maximum of the dichroic effect in the weak-laser-field domain
is found from Eqs.  (\ref{dcsCDe}) and (\ref{RCDADe})
for  an optimal laser intensity ratio $I_3/I_1 \simeq 20.2$, where $a_1 \simeq a_2$
and
$R_{CDAD}( \theta,\varphi ) \simeq
 \sin (\varphi) \sin (3\varphi)/
[1+   \cos (\varphi) \cos (3\varphi)]$.
Therefore, the advantage of analyzing the relative difference between the angular distributions obtained for two-color co- and counterrotating CP fields, Eq.  (\ref{RCDADe}),
is that we can determine the relative magnitude, $a_1/a_2$, of the interfering transition amplitudes.

\subsection{Two-photon circular dichroism at moderate laser intensity}
\label{IVb}

\begin{figure}
\centering
\includegraphics[width=4.5in,angle=0]{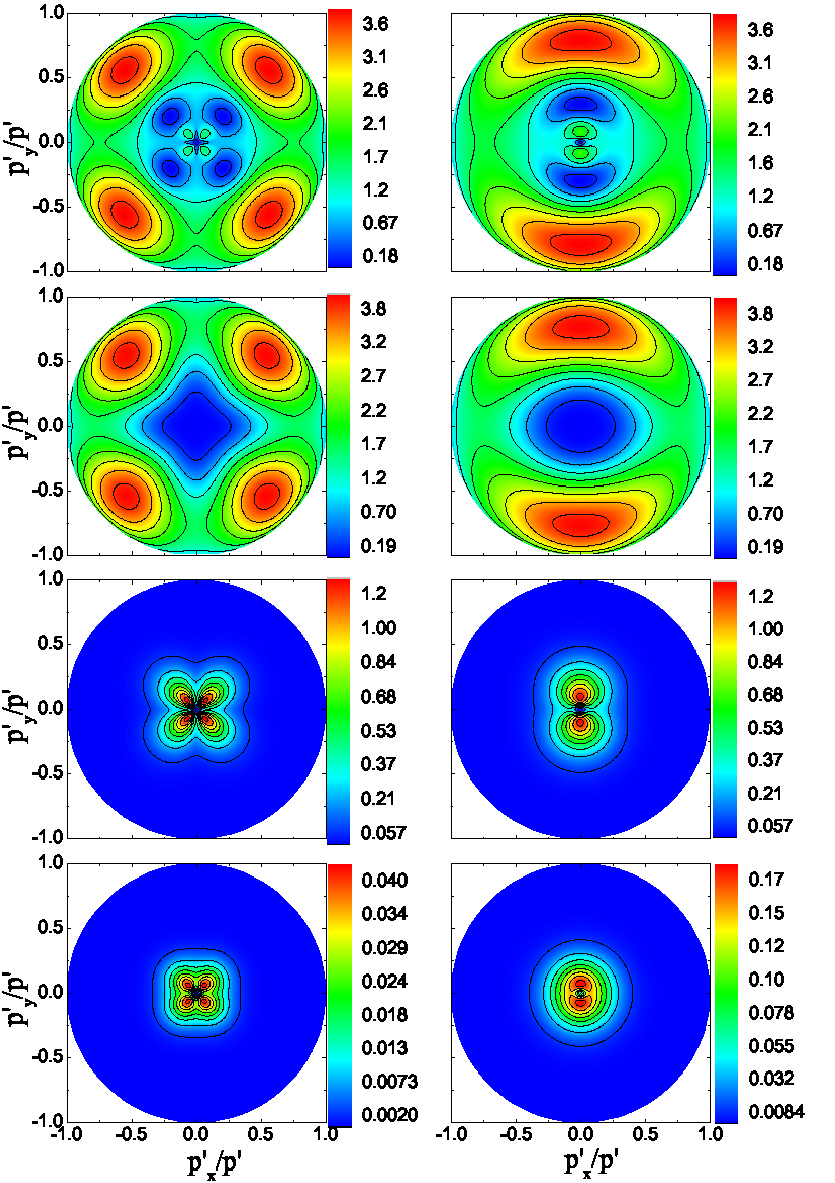}
\caption{(Color online)
Contour plots representing  DCSs ($N=2$),
given by Eq. (\ref{dcs}),  for  two-color left-handed CP fields in the  right column
and two-color left- and right-handed CP fields in the left column,
  as a function of the normalized  Cartesian  components of the projectile momentum vector
 in the polarization plane, $p_{x}^\prime/{p^\prime} $ and $p_{y}^\prime/{p^\prime} $.
The plots from top to bottom represent the total DCS,
the projectile electron contribution calculated as $(2\pi)^4 (p'/p )|T_2^{(0)}|^2$,
the first-order atomic dressing contribution,
 $(2\pi)^4 (p'/p )|T_2^{(1)}|^2$,
and  the second-order atomic dressing contribution
$(2\pi)^4 (p'/p )|T_2^{(2)}|^2$.
The projectile electron energy is $ E_{p}=100$ eV, $\textbf{p} \parallel \mathbf{e}_{z}$,
 the laser field intensities are $I_1=I_3=1$ TW/cm$^2$,
 and the photon energies are $\omega_1= 3 $ eV and $\omega_3=3\omega_1$.
The DCSs in a.u. are multiplied by a $10^{4}$ factor and
their magnitudes are indicated by the color scales in each row.
}
\label{fig4}
\end{figure}

Now, we apply the semiperturbative formulas derived in Sec. \ref{II},
 Eqs. (\ref{tn0}), (\ref{tn1}), and (\ref{t2}), to evaluate numerically the
nonlinear DCSs and CDAD for two-photon absorption ($N = 2$) in  elastic  electron scattering
by a hydrogen atom  in the presence of  two-color co- and counterrotating CP laser fields.
We  have chosen moderate laser intensities below $10$ TW/cm$^2$,
 a high energy of the projectile electron $ E_{p}=100$ eV, and
  photon energies in the UV range $\omega_1=3$ eV and $\omega_3= 9 $ eV,
 such that neither the projectile electron nor  the photon can separately
excite an upper atomic state.
The polarization vectors of the two-color bicircular laser field are given by
$ \bm{\varepsilon}_{1}  = (\mathbf{e}_x+i \mathbf{e}_y)/\sqrt{2}$  and
 $ \bm{\varepsilon}_{3}  =\bm{\varepsilon}_{1} $ for the corotating case, and
$ \bm{\varepsilon}_{1}  = (\mathbf{e}_x+i \mathbf{e}_y)/\sqrt{2}$ and
$ \bm{\varepsilon}_{3}  =(\mathbf{e}_x-i \mathbf{e}_y)/\sqrt{2} $ for the counterrotating case.
\noindent
To start with a simple case, we present in Fig. \ref{fig4}  our numerical results
for the three-dimensional  DCSs, projected in the polarization plane,
as a function of the normalized projectile momentum
$p_{x}^\prime/{p^\prime} $ and $p_{y}^\prime/{p^\prime} $
 for (a) two-color left-handed CP fields (corotating)  in the right column
 and  (b) left- and right-handed CP fields (counterrotating) in the left column.
The  intensities of the fundamental and third-harmonic laser are considered equal, as
$I_1=I_3=1$ TW/cm$^2$, which result in a quiver motion amplitude
$\alpha_{01} \simeq 0.44$ a.u. and an argument of the Bessel function
${\cal R}_1 \simeq  0.44 |\bm{\varepsilon}_1\cdot {\mathbf{q}}|$ for the fundamental field,
while for the harmonic field the corresponding parameters
 $\alpha_{03} $ and ${\cal R}_3$ are 9 times smaller.
At laser intensities higher than 10 TW/cm$^2$ the multiphoton ionization  becomes a competing process and should be taken into account.
 The total DCS,  the projectile contribution to DCS calculated as
 $(2\pi)^4 (p'/p )|T_2^{(0)}|^2$, Eq. (\ref{tn0}),
and  the first- and second-order atomic dressing contributions calculated as
 $(2\pi)^4 (p'/p )|T_2^{(1)}|^2$ and $(2\pi)^4 (p'/p )|T_2^{(2)}|^2$,
Eqs. (\ref{tn1}) and (\ref{t2}),  are shown from top to bottom in Fig. \ref{fig4}.
The first- and second-order atomic contributions to DCSs, i.e., the last two rows in Fig. \ref{fig4},
give significant contributions at relatively small scattering angles $\theta <30^\circ$,
 while the projectile electron contribution, i.e., the second row in Fig. \ref{fig4},
gives important contributions at larger scattering angles.
It is clear that, at small scattering angles  the differences between
the co- and counterrotating DCSs originate
 from the different first- and second-order atomic dressing,
whereas at large scattering angles the differences come from the different
projectile electron contributions to DCSs.
Our numerical results in Fig. \ref{fig4} show a strong dependence of the DCSs on the
scattering and azimuthal angles, as well on the photon  helicities of the two bicircular fields.
For counterrotating CP fields DCS is invariant  to rotation around the $z$ axis
 by an   $\varphi=\pi/2$ azimuthal angle, whereas
for corotating CP fields the corresponding  azimuthal angle  is $ \varphi=\pi$.
Both DCSs  for co- and counterrotating polarizations
are symmetric with respect to reflection in the $(x,z)$- and $(y,z)$-planes, such that
$ {d{\sigma}_2}( \theta, \pi-\varphi )/{d\Omega_{p'}}
 ={d{\sigma}_2}( \theta, \pi+\varphi )/{d\Omega_{p'}}$
and
$ {d{\sigma}_2}( \theta, \pi/2-\varphi )/{d\Omega_{p'}}
 ={d{\sigma}_2}( \theta, \pi/2+\varphi )/{d\Omega_{p'}} $, respectively.
Recently, similar rotational and reflection symmetries were obtained in the
differential ionization rate in above-threshold ionization of krypton atoms
 by a two-color bicircular laser field of $\omega$ and  $3\omega$ frequencies \cite{Hasovic}.

\begin{figure}
\centering
\includegraphics[width=4.5in,angle=0]{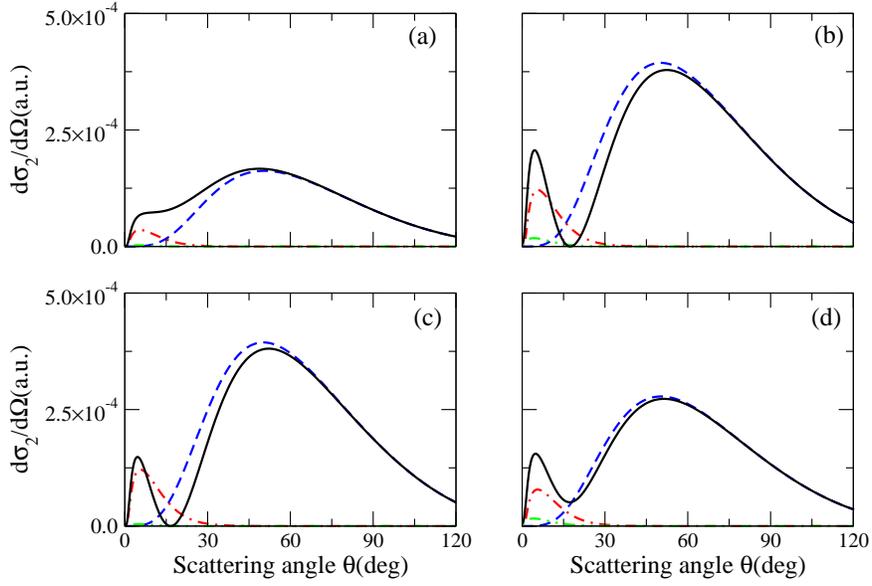}
\caption{(Color online)
The total two-photon DCSs (full lines) by two-color left- and right-handed-CP
 laser fields in panels (a) and (c)
and two-color left-handed-CP  laser fields  in panels (b) and (d)
as a function of the scattering angle  $\theta$. The azimuthal angles are
$\varphi=90^\circ$ in panels (a) and (b), and $\varphi=45^\circ$ in  panels (c) and (d),
and the rest of the parameters are the same as in  Fig. {\ref{fig4}}.
The dashed lines represent the projectile electron contribution
calculated as $(2\pi)^4 (p'/p )|T_2^{(0)}|^2$,
while the dot-dashed  and  dotted  lines represent
the first- and second-order atomic dressing contributions,
 $(2\pi)^4 (p'/p )|T_2^{(1)}|^2$
and $(2\pi)^4 (p'/p )|T_2^{(2)}|^2$, respectively.
}
\label{fig5}
\end{figure}

For a better understanding of the contour plots presented in Fig. \ref{fig4},
 we show in Fig. \ref{fig5}  the DCSs as a function of the  scattering angle, $\theta$,
for two-color left-handed CP fields (corotating case)  in the right column
and for left- and right-handed CP fields (counterrotating case) in the left column.
The DCSs are plotted at two azimuthal angles $\varphi = 90^{\circ} $
in Figs. \ref{fig5}(a) and \ref{fig5}(b), and $\varphi =45^{\circ} $
 in  Figs. \ref{fig5}(c) and \ref{fig5}(d),  while the other parameters
concerning the scattering geometry, the field intensities, and
the projectile and photon energies  are the same as in  Fig. \ref{fig4}.
The dashed lines represent the projectile electron contribution to DCS
calculated as $(2\pi)^4 (p'/p )|T_2^{(0)}|^2$,
while the dot-dashed  and  dotted  lines represent
the first- and second-order atomic dressing contributions given by
 $(2\pi)^4 (p'/p )|T_2^{(1)}|^2$ and $(2\pi)^4 (p'/p )|T_2^{(2)}|^2$, respectively.
As expected from our theoretical calculations, the first-order laser-atom interaction
 is quite important at small scattering angles $\theta<20^\circ $.
By contrast, at larger scattering angles ($\theta >20^\circ$)
the  projectile electron contribution to DCS is dominant
due to nuclear scattering and determines the angular distribution of the two-photon DCS,
as is shown by the dashed lines in Fig. \ref{fig5}.
The projectile electron  is scattered with the highest probability
at the scattering angles $\theta \simeq 50^\circ $
in Figs. \ref{fig5}(a) and \ref{fig5}(c),
and  $\theta \simeq 52^\circ $  in Figs. \ref{fig5}(b) and \ref{fig5}(d).

\begin{figure}
\centering
\includegraphics[width=4.5in,angle=0]{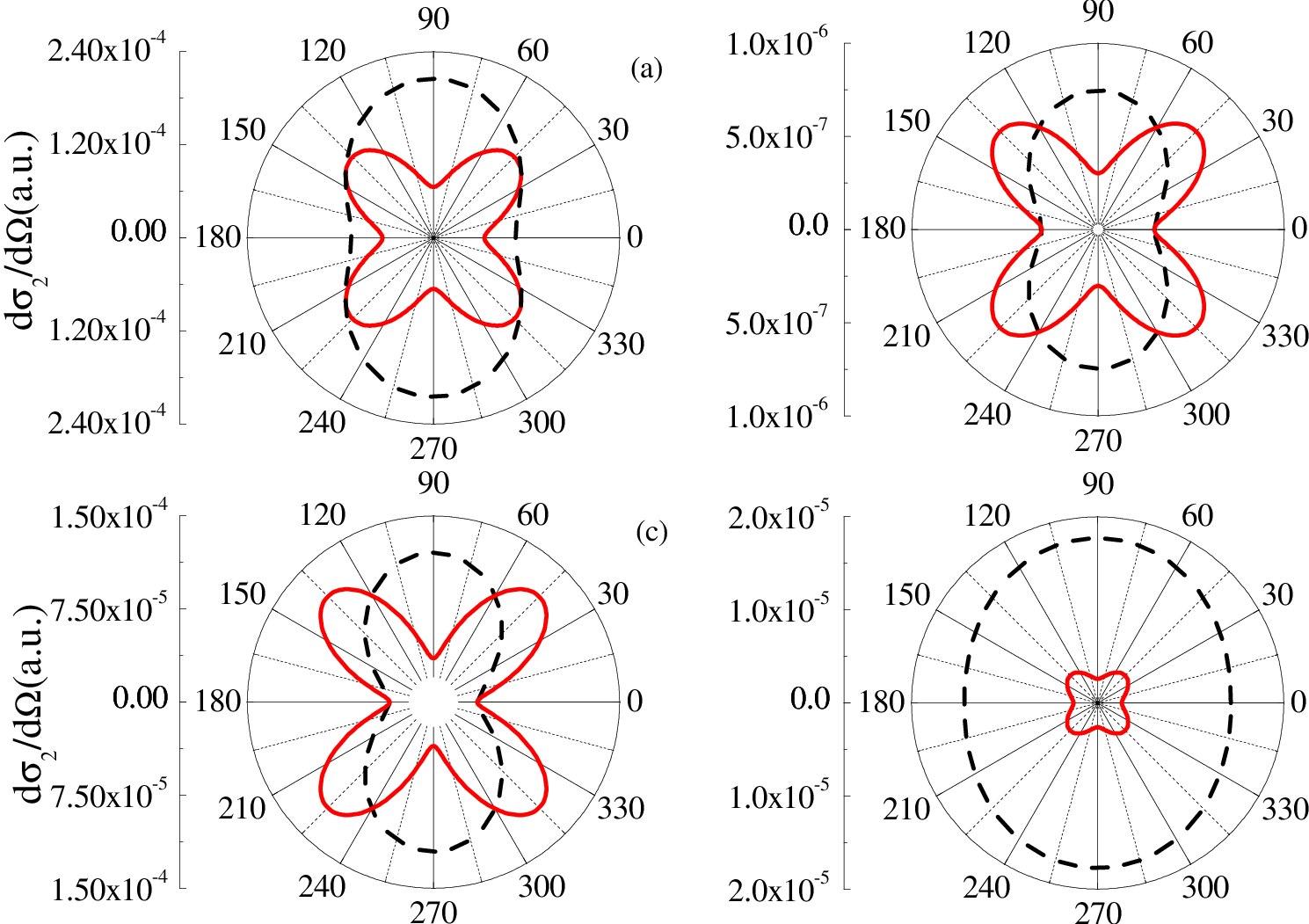}
\caption{(Color online)
Two-photon DCSs by  two-color left- and  right-handed-CP (full lines) laser fields
and  two-color left-handed-CP (dashed lines) laser fields are shown in panel (a)
as a function of the  azimuthal angle $\varphi$.
The projectile electron contribution
calculated as $(2\pi)^4 (p'/p )|T_2^{(0)}|^2$ is plotted in panel (b),
the first-order atomic dressing contribution,
 $(2\pi)^4 (p'/p )|T_2^{(1)}|^2$ is in panel (c),
and  the second-order atomic dressing contribution
$(2\pi)^4 (p'/p )|T_2^{(2)}|^2$ is in panel (d).
The scattering angle is fixed at $\theta=5^\circ$ and
the rest of the parameters are the same as in  Fig. {\ref{fig4}}.
}
\label{fig6}
\end{figure}

In order clarify the importance of the atomic dressing effect on the scattering signal
presented in Fig. \ref{fig4},  we illustrate  in Fig. \ref{fig6}(a) the two-photon DCSs in polar plots
for counterrotating (solid lines) and corotating (dashed lines) CP fields
 as a function of the azimuthal angle, $\varphi$,
 at the scattering angle $\theta = 5^\circ $, while
the rest of the parameters are the same as in Fig. \ref{fig4}.
The projectile  contribution to DCS, $(2\pi)^4 (p'/p )|T_2^{(0)}|^2$, is plotted in Fig. \ref{fig6}(b),
while the first- and second-order atomic dressing contributions, calculated as
 $(2\pi)^4 (p'/p )|T_2^{(1)}|^2$ and $(2\pi)^4 (p'/p )|T_2^{(2)}|^2$, are depicted
in Figs. \ref{fig6}(c) and  \ref{fig6}(d), respectively.
Because of the strong first- and second-order atomic dressing effects
at small scattering angles the atomic contribution is about 2 orders of magnitude
 larger compared to the  projectile electron contribution.
Moreover, at small scattering angles there is a clear enhancement of  DCSs for
corotating compared to counterrotating CP fields and, obviously,
the differences in magnitude between the  DCSs for co- and counterrotating polarizations
 originate  from the different second-order atomic dressing terms, as is shown in
the last row of Fig. \ref{fig4}  and   Fig. \ref{fig6}(d).
The two-photon DCS for corotating CP fields has a  pattern profile predicted
in the weak-laser-field regime by the $e^{-2i\varphi}$  term in Eq. (\ref{dcsLL})
and the projectile electron is scattered with a  high probability in
the azimuthal angle directions $ \varphi=\pi/2$ and  $3\pi/2$.
By contrast, the two-photon DCS for counterrotating CP fields, as well as
the electronic, first- and second-order atomic contributions
 to DCS have a specific \textquotedblleft four-leaf clover\textquotedblright pattern,
 as is analytically predicted in the weak-laser-field regime by the $e^{-4 i\varphi}$ term  in Eq. (\ref{dcsLR}),
and the projectile electron is scattered with a  high probability in the azimuthal directions
$ \varphi=\pi/4, 3\pi/4,5\pi/4$, and  $7\pi/4$, as presented in Fig. \ref{fig4}.
The  two-photon DCS is invariant  to rotation around the $z$ axis by an azimuthal angle
$ \varphi=\pi$ for  corotating CP fields and   $ \varphi=\pi/2$ for counterrotating CP fields,
and DCSs are symmetric with respect to reflection in the $(x,z)$- and  $(y,z)$-planes
for both co- and counterrotating CP fields.

\begin{figure}
\centering
\includegraphics[width=4.5in,angle=0]{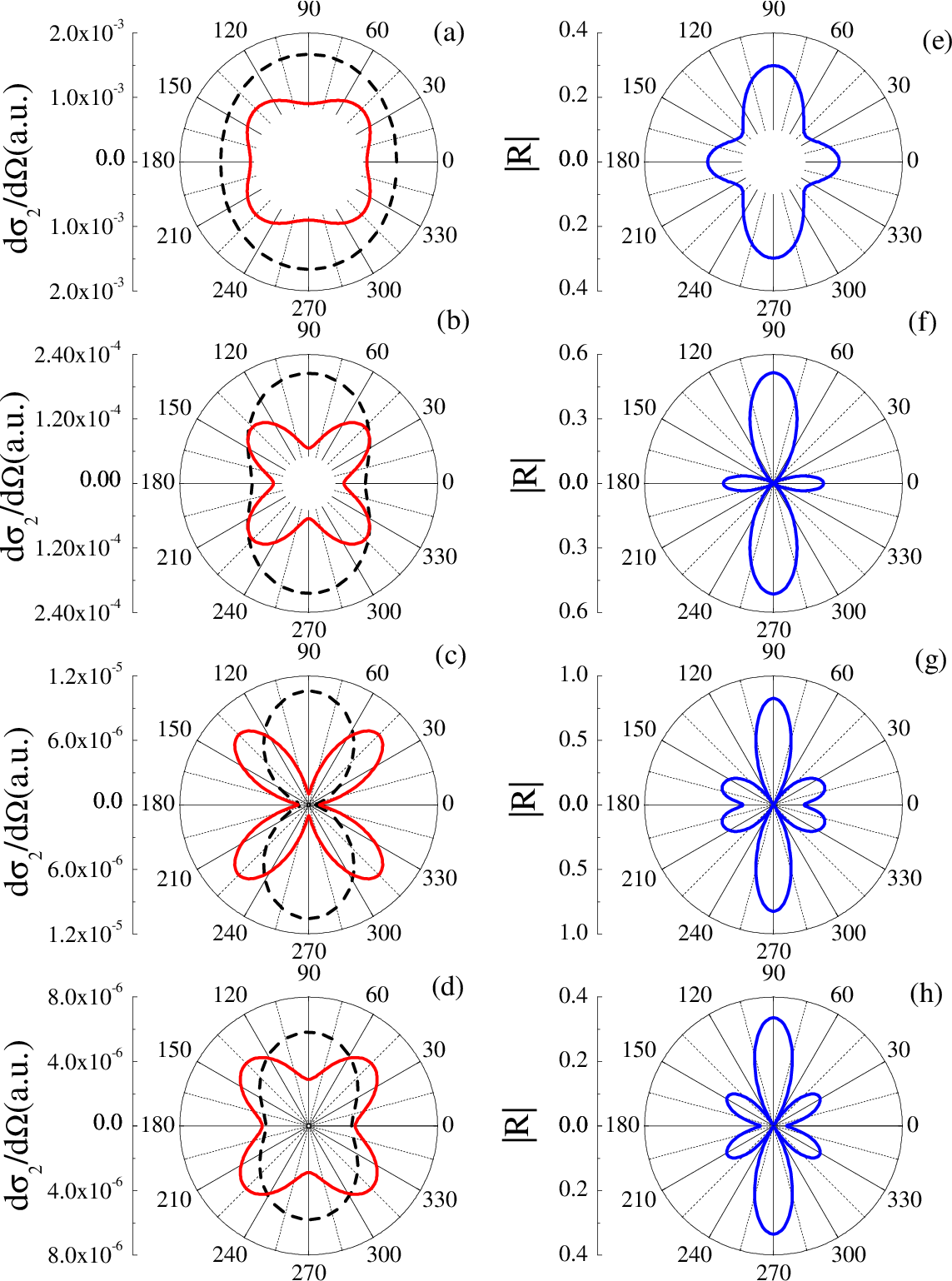}
\caption{(Color online)
The total two-photon DCSs   in the left column
and the absolute value of the relative CDAD, $|R(\varphi)|$, in the right column
by  two-color left- and right-handed-CP (full lines) laser fields
and  two-color left-handed-CP (dashed lines) laser fields,
as a function of the  azimuthal angle $\varphi$.
The scattering angle is fixed at $\theta=5^\circ$, and intensities of the harmonic field
$I_3$ are equal to $10I_1$ in panels (a) and (e), $I_1$ in panels (b) and (f),
$10^{-2}I_1$ in panels (c) and (g),  and $10^{-3}I_1$ in panels (d) and (h).
 The rest of the parameters are the same as in  Fig. {\ref{fig4}}.
}
\label{fig7}
\end{figure}

 Finally, we present the  DCSs  for two-photon absorption
in Figs. \ref{fig7}(a)-\ref{fig7}(d) for co-and counterrotating polarizations
and the absolute values of the relative CDAD, $|R(\varphi)|$,
 given by Eq. (\ref{RCDAD})  in Figs. \ref{fig7}(e)-\ref{fig7}(h)
as a function of the azimuthal angle of the scattered electron.
The  scattering angle is fixed at the value $\theta=5^\circ$ and
 the intensity of the third-harmonic laser is  $I_3= f I_1$,
 with the laser intensity ratios $f = 10,1,10^{-2}$, and $10^{-3}$ from top to bottom
in Fig. \ref{fig7}, while the rest of the parameters are the same as in Fig. \ref{fig4}.
For the harmonic field these laser  parameters correspond to a quiver motion amplitude
$\alpha_{03} = \alpha_{01}\sqrt{f} /9$ and an argument of the Bessel function
${\cal R}_3 = {\cal R}_1 \sqrt{f} /9$.
At small scattering angles  and relatively low intensities of the harmonic field,
Figs. \ref{fig7}(b)-\ref{fig7}(d),
 there is a strong dependence  of DCSs on the azimuthal angle due to
the first- and second-order atomic dressing terms for  both co- and counterrotating CP lasers,
while at a higher intensity, $I_{3} =10I_1$ plotted in Figs. \ref{fig7}(a) and \ref{fig7}(e),
 there is a weaker dependence of DCSs on the azimuthal angle
because of the dominance of the two different photon processes shown in Fig. \ref{fig3}(b).
The absolute values of relative CDAD, $|R(\varphi)|$,  are larger at the azimuthal angles
  $\varphi=\pi/2$ and $3\pi/2$ due to  the different
two- and four-fold symmetries of DCSs for co- and counterrotating CP fields,
as shown  in  Figs. \ref{fig7}(a)-\ref{fig7}(d) and Figs. \ref{fig7}(e)-\ref{fig7}(h).
Therefore, at small scattering angles, because of the strong first- and second-order atomic dressing effects, there is a clear enhancement of  DCSs for corotating compared to counterrotating CP fields,
as shown in Figs. \ref{fig4} and \ref{fig6}, as well as an important dichroic effect.

\begin{figure}
\centering
\includegraphics[width=4in,angle=0]{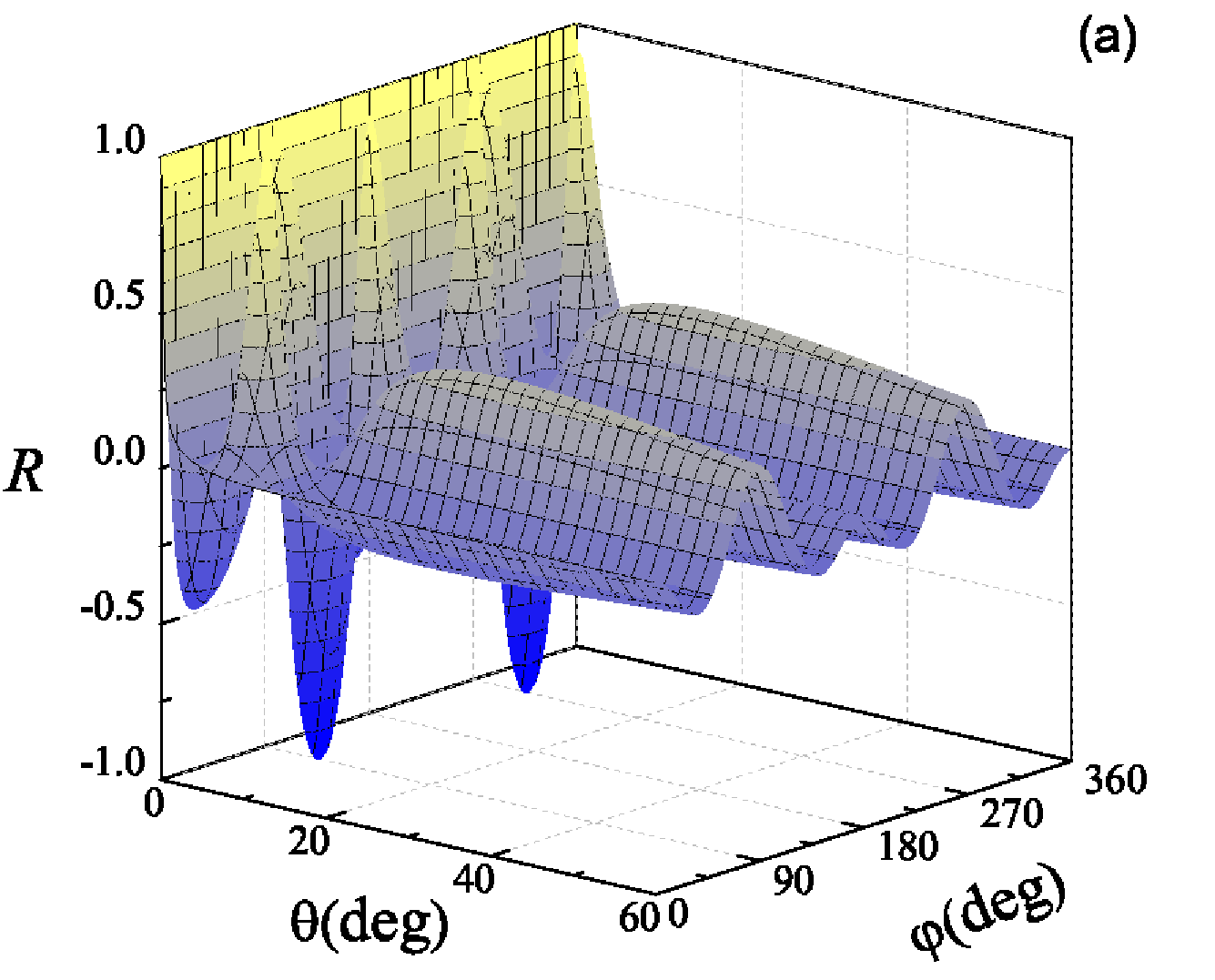}
\includegraphics[width=4in,angle=0]{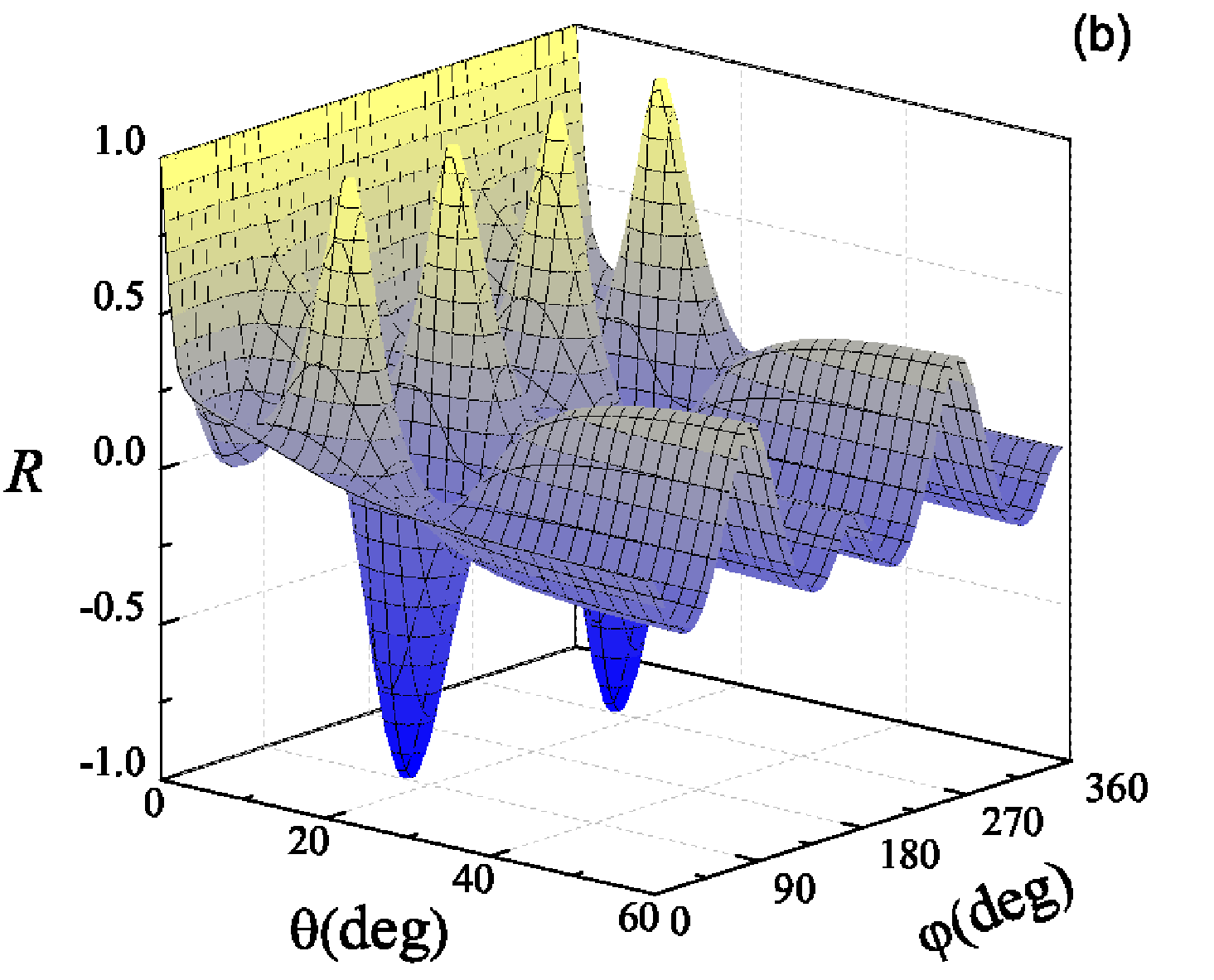}
\caption{(Color online)
Relative CDAD, $R(\theta,\varphi)$,  as a function of the scattering and azimuthal angles,
for photon energies $\omega_{1} = 1.5 $ eV  in panel (a)  and $\omega_{1} = 3 $ eV in panel  (b).
 The rest of the parameters are the same as in  Fig. {\ref{fig4}}.
}
\label{fig8}
\end{figure}

\begin{figure}
\centering
\includegraphics[width=3.0in,angle=0]{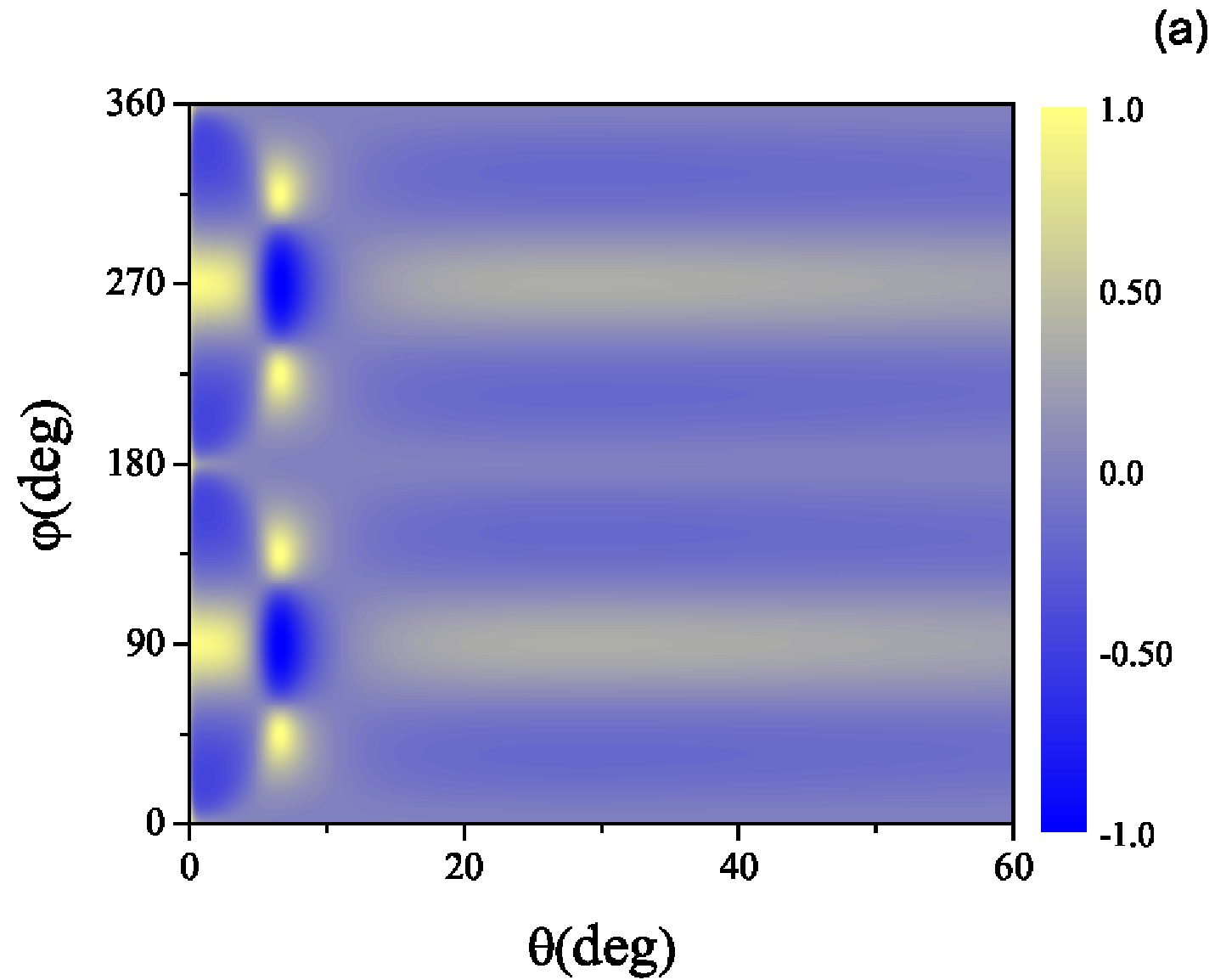}
\includegraphics[width=3.0in,angle=0]{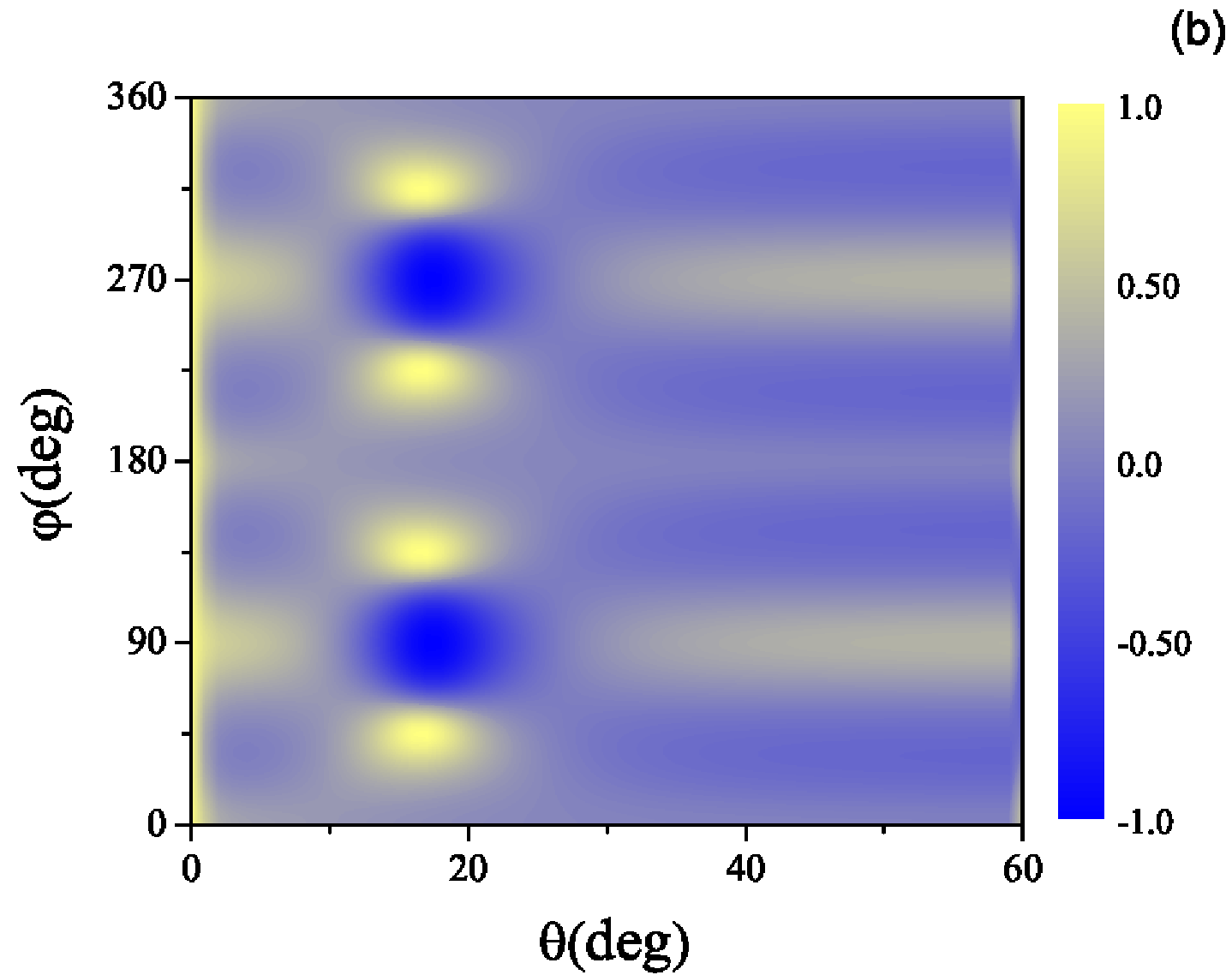}
\caption{(Color online)
Contour plots presenting the same results as in Fig. {\ref{fig8}},
namely relative CDAD, $R(\theta,\phi)$,  as a function of the scattering and
azimuthal angles, $\theta$ and $\varphi$,
 for $\omega_{1} = 1.5 $ eV  in panel (a)  and $\omega_{1} = 3$ eV in panel  (b).}
\label{fig9}
\end{figure}

\noindent
We emphasize that the  dichroic effect in DCS for two-photon absorption is not,
 however, a one-photon resonance effect
and it strongly depends on the atomic dressing by the bicircular laser field.
We demonstrate this by comparing  in Figs. \ref{fig8}(a) and \ref{fig8}(b)
 the three-dimensional numerical data
for the relative values of the CDAD, $R(\theta,\varphi)$, calculated from Eq. (\ref{RCDAD})
as a function of the scattering and azimuthal angles
at two non-resonant photon energies $\omega_1 =1.5$ eV and $\omega_1 =3$ eV.
The other parameters are the same as in Fig. \ref{fig4}, namely  $E_p=100$ eV and
  $I_1=  I_3=1$ TW/cm$^2$.
For a clear view of the positive and negative values of relative CDAD
the same results of  Figs. \ref{fig8}(a) and \ref{fig8}(b) are plotted as contour plots
in  Figs. \ref{fig9}(a) and \ref{fig9}(b).
At $\omega_1 =1.5 $ eV  the dichroic effect in DCS is as well very important
but at relatively small scattering angles only,  i.e., $\theta<10^{\circ}$,
 where the atomic dressing dominates.
By contrast, at $\omega_1 =3 $ eV  there is a stronger dependence of the relative CDAD
on both scattering and azimuthal angles because of the atomic
dressing effects that occur at larger photon energies \cite{acgabi99}.
The dichroic effect is quite important at relatively small scattering angles
 $\theta<25^{\circ}$ and is not negligible  even at larger scattering angles.
At small scattering angles since the DCSs
take very small values frequently due to the destructive interference between the
projectile and atomic contributions,
 as depicted by the blue areas in the first row of Fig. \ref{fig4},
the relative CDAD oscillates quite rapidly between  $-1$ and $+1$
in Figs. \ref{fig8} and \ref{fig9}.
Furthermore, as predicted from the closed formulas of DCSs
 for co- and counterrotating CP fields,
${d\sigma_{N}^{++}}/{d\Omega_{p'}}$ or $ {d\sigma_{N}^{+-}}/{d\Omega_{p'}}$
 given by Eqs.  (\ref{dcsLL}) and (\ref{dcsLR})
and Eqs.  (\ref{c1l})-(\ref{c3l}),
 at moderate laser intensities below $ 1$ TW/cm$^2$ and larger scattering angles
 where the atomic dressing becomes negligibly small, namely
 $\theta >15^{\circ}$ for $\omega_1 =1.5 $ eV
and $\theta >40^{\circ}$ for $\omega_1 =3 $ eV,  the dependence
 of the relative dichroism on the scattering and azimuthal angles in
Figs. \ref{fig8} and \ref{fig9}  is well described by Eq. (\ref{RCDADe}).

\section{Summary and conclusions}
\label{IV}

Using a semiperturbative method, we have theoretically studied the dichroic effect in
electron-hydrogen scattering  by a two-color bicircular laser fields
 of commensurate frequencies and moderate intensities.
We have investigated a different regime of CD where the monochromatic components
of the two-color CP field rotate in the same plane with
the same or opposite helicities.
We  predict the existence of a nonlinear dichroic effect in DCS
 at high scattering projectile energies,
which is sensitive to the photon energies and laser field intensities.
We have derived useful analytical formulas for two-photon CDAD,
 that give more physical insight into the  scattering process
and valuable information for experimental investigations.
We stress that the analytical formulas obtained  for co- and counterrotating
  polarizations  in the weak-laser-field limit,  Eqs.  (\ref{dcsLL}) and (\ref{dcsLR}),
indicate that the two-photon DCSs are  related to the interference of different
quantum paths involving two photons with identical or different polarizations.
By varying the intensity ratio of the   co- and counterrotating two-color CP laser field
components  we can manipulate the angular distribution of the scattered electrons.
We have established that at UV photon energies and small scattering angles
there is a clear enhancement of the DCS
 for corotating compared to counterrotating  laser fields,
 because of the strong second-order atomic dressing effects.
The dichroic effect in the angular distribution of scattered electrons
 originates from the nonzero atomic dressing at small scattering angles,
 whereas at large scattering angles the dichroic effect comes from the
projectile contribution to the scattering signal.
\noindent
In conclusion, the investigation of CDAD in the scattering signal is an effective
method of studying the polarization effects in laser-assisted electron-atom collisions
and we hope that the dichroic effect discussed in the present paper
will be useful in future theoretical and experimental studies.

\newpage

\section*{ACKNOWLEDGMENTS}
The work by G. B. was supported by  research programs PN 16 47 02 02 through
Contract No. 4N/2018 (Laplas V) and  FAIR-RO Contract No. 01-FAIR/2016
from the UEFISCDI and the Ministry of Research and Innovation of Romania.

\clearpage
\newpage

\newpage

\newpage

\end{document}